\documentclass[journal]{IEEEtran}

\IEEEoverridecommandlockouts

\usepackage{graphics} 
\usepackage{epsfig} 
\usepackage{mathptmx} 
\usepackage{times} 
\usepackage{amsmath} 
\usepackage{amssymb}  
\usepackage{multirow}
\usepackage{epstopdf} 
\usepackage{cite} 
\usepackage{stfloats} 
\usepackage{bm}
\usepackage{color}
\usepackage{amsthm}
\usepackage{cite}
\usepackage{array}
\newtheorem{remark}{Remark}
\usepackage{threeparttable}
\usepackage{nomencl}
\usepackage{braket}
\usepackage[switch,pagewise,columnwise]{lineno}
\usepackage{algorithm}
\usepackage{algorithmic}
\usepackage{caption}

\usepackage{booktabs} 
\usepackage{multirow} 
\usepackage{makecell} 
\usepackage{threeparttable}

\makenomenclature

\IEEEoverridecommandlockouts    

\newcounter{algsubstate}

\title{Learning Informative Latent Representation for Quantum State Tomography}

\author{Hailan~Ma, Zhenhong~Sun, Daoyi~Dong, Dong Gong
	\thanks{This work was supported by the Australian Research Council’s Future
		Fellowship funding scheme under Project FT220100656 to D. Dong and an ARC DECRA Fellowship DE230101591 to D. Gong. (Corresponding author: Daoyi Dong).}%
	\thanks{H. Ma is with the School of Engineering 
		and Technology, University of New South Wales, Canberra, ACT 2600,
		Australia and is also with the School of  Engineering, The Australian National University, Canberra, ACT 2601, Australia (email: hailanma0413@gmail.com).}
	\thanks{Z. Sun and D. Dong are with the School of  Engineering, The Australian National University, Canberra, ACT 2601, Australia (email: zhenhong.sun@anu.edu.au, daoyi.dong@anu.edu.au)}
	\thanks{D. Gong is  with the School of Computer Science and Engineering, University of New South Wales, NSW 2052, Australia (email: dong.gong@unsw.edu.au)}
}

\begin{document}
	\maketitle

	\begin{abstract}
		
Quantum state tomography (QST) is the process of reconstructing the complete state of a quantum system (mathematically described as a density matrix) through a series of different measurements. These measurements are performed on a number of identical copies of the quantum system, with outcomes gathered as probabilities/frequencies. 
QST aims to recover the density matrix and the corresponding properties of the quantum state from the measured frequencies. 
Although an informationally complete set of measurements can specify the quantum state accurately in an ideal scenario with a large number of identical copies, both the measurements and identical copies are restricted and imperfect in practical scenarios, making QST highly ill-posed. The conventional QST methods usually assume adequate or accurate measured frequencies or rely on manually designed regularizers to handle the ill-posed reconstruction problem, suffering from limited applications in realistic scenarios. 
Recent advances in deep neural networks (DNNs) led to the emergence of deep learning (DL) in QST. However, existing DL-based QST approaches often employ generic DNN models that are not optimized for imperfect conditions of QST. In this paper, we propose a transformer-based autoencoder architecture tailored for QST with imperfect measurement data. Our method leverages a transformer-based encoder to extract \emph{an informative latent representation} (ILR) from imperfect measurement data and employs a decoder to predict the quantum states based on the ILR. We anticipate that the high-dimensional ILR will capture more comprehensive information about the quantum states. To achieve this, we conduct pre-training of the encoder using a pretext task that involves reconstructing high-quality frequencies from measured frequencies. Extensive simulations and experiments demonstrate the remarkable ability of the informative latent representation to deal with imperfect measurement data in QST.
	
	\end{abstract}
	
	\begin{IEEEkeywords}
	quantum state tomography, imperfect measurement data, fidelity, latent representation, pre-training. 
	\end{IEEEkeywords}

	\section{Introduction}\label{sect:intro}

\begin{figure*}[t]
		\centering
		\begin{minipage}[b]{0.48\textwidth}
			\centering
			\includegraphics[width=0.9\textwidth]{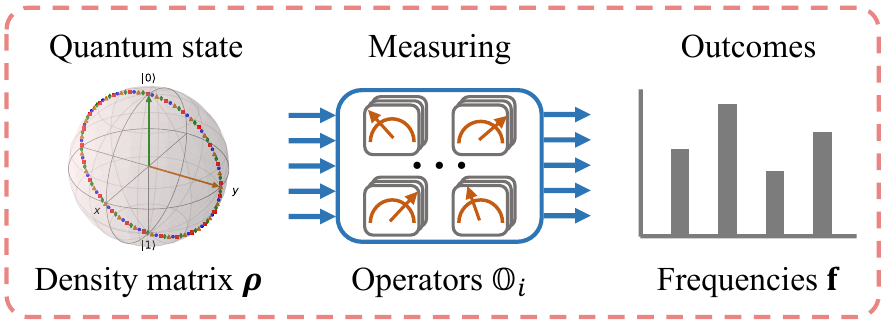}
			\centerline{(a) State measurement}
			\label{sfig:nnqst1}
		\end{minipage}
		\begin{minipage}[b]{0.48\textwidth}
			\centering
			\includegraphics[width=0.9\textwidth]{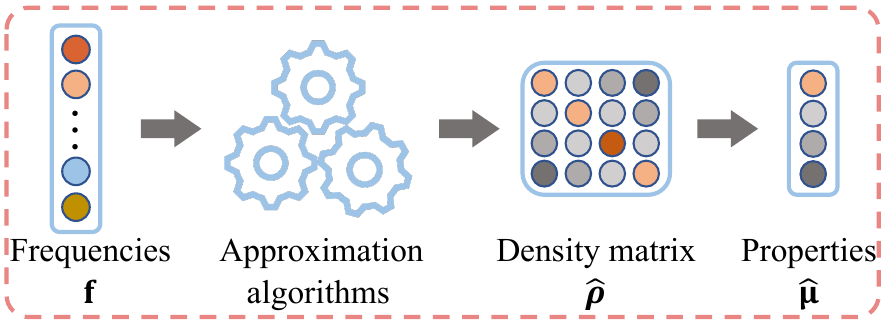}
			\centerline{(b) State reconstruction}
			\label{sfig:nnqst2}
		\end{minipage}
		\caption{Schematic of QST.  (a) Perform different measurements on quantum states and obtain frequencies from collected outcomes. (b) Infer quantum representations, including density matrix (full description) and quantum properties (partial description) from measured frequencies.}
		\label{fig:nnqst}
	\end{figure*}

	The determination of the quantum state of a system, known as quantum state tomography (QST), is the gold standard for verification and benchmarking of quantum devices and is also essential for engineering systems in developing quantum technology~\cite{gebhart2023learning,rambach2021robust}. In the physical world, the full description of a quantum state can be represented as a density matrix $\rho$, i.e., a positive-definite Hermitian matrix with unit trace. To determine $\rho$, one may first perform measurements on a collection of identically prepared copies of a quantum system, gathering statistical outcomes to a set of frequencies $\mathbf{f}$, as shown in Fig.~\ref{fig:nnqst}(a).
	Then, the reconstruction of quantum states is realized by mapping the measured frequencies $\mathbf{f}$ into a full description of a quantum state (density matrix $\rho$) or a partial description of a quantum state (quantum properties $\bm{\mu}$) using approximation algorithms, as illustrated in  Fig.~\ref{fig:nnqst}(b).
	
Generally, to uniquely identify a quantum state, the measurements must be informatively complete to provide all the information about $\rho$~\cite{jevzek2003quantum}, except for some special cases when density matrices of quantum states are low-rank~\cite{haah2017sample} or take the form of matrix product operators~\cite{qin2023stable}. The exponential scaling of parameters in $\rho$ requires an exponentially increasing number of measurements, each of which requires a sufficient number of identical copies~\cite{gebhart2023learning}, posing a great challenge in practical applications. For example, in solid-state systems, the process associated with measuring one copy of a quantum state can be time-consuming, and implementing a sufficient number of measurement operators requires complex and costly experimental setups. Given a coherence time (beyond which quantum states might change), the total copies of identical states for measurements may be constrained. In this scenario, the measurement data may be a complete but inaccurate frequency vector (a few copies assigned to each measurement operator ) or an accurate but incomplete frequency vector~\cite{len2022quantum} (sufficient copies assigned to partial measurement operators that are experimentally easy to generate~\cite{chantasri2019quantum}). The two factors are collectively referred to as imperfect scenarios (i.e., ill-posed problems) in this paper, which hinder the precise reconstruction of a quantum state. 

	When faced with imperfect statistics obtained in practical scenarios, traditional methods either treat those imperfections as noise in accurately measured frequencies, without employing a specific methodology~\cite{jevzek2003quantum} or require manually-designed regularizers~\cite{teo2012incomplete} to handle the ill-posed problem. Recently, neural network (NN) approaches have demonstrated their generality and robustness in learning intricate features from quantum systems~\cite{gebhart2023learning}, including restricted Boltzmann machine (RBM)~\cite{torlai2018neural} and recurrent neural network (RNN)~\cite{carrasquilla2019reconstructing,cha2021attention} and generative adversarial network (GAN)~\cite{ahmed2021quantum} for QST tasks. However, these methods have not yet incorporated imperfect factors into their network design and thus exhibit sub-optimal performance when confronted with imperfect measurement data. Practical QST suffers from noise addition and information loss in the measured frequencies, reducing the representation of measured data for accurate quantum state reconstruction. Although the capability of neural networks in dealing with incomplete measurement has been investigated in~\cite{danaci2021machine}, their method relies on a separate statistical technique to infer missing values from the rest of the data and involves the decoupling of various functionalities gathered in one single model into different models.
	With the emergence of new architectures of NNs designed for practical problems~\cite{transformer2017,devlin2018bert}, a natural question arises: is it possible to develop an enhanced architecture that corporates imperfect measurement data to learn an informative representation from imperfect measurement data?

	Recognizing the representation reduction posed by the ill-posed problem, we turn to a powerful neural architecture to extract a latent representation that might compensate for imperfect data. Specifically, we propose a transformer-based autoencoder to address the challenge of imperfect measurement data in practical QST applications.
	The encoder serves as an extractor to capture the informative latent representation (ILR) of quantum states from measured data. Through the state decoder, the latent representation then provides a useful feature that can be further leveraged to estimate a quantum state, including its density matrix and significant properties. In typical ML tasks, supervised training, by its nature, is a guided process. When it relies solely on imperfect data as the ground truth, it restricts the intermediate layers from fully realizing their potential, thereby limiting the richness of the latent representations. Hence, simply training the autoencoder model with raw measured data might not entirely capitalize on the potential of intermediate layers to abstract meaningful latent representations from the imperfect data. To overcome the limitation of imperfect measured data, we propose augmenting the encoder with a frequency decoder during pre-training, which helps to extract a highly informative latent representation.

	
	To ensure that the encoder's latent representation is sufficiently informative while maintaining model expressiveness predominantly in the encoder, the pre-training autoencoder is designed with an asymmetrical architecture, where the encoder is larger in scale compared to the decoder. 
	The entire design facilitates the acquisition of an informative latent representation from imperfect measurement data, which is highly beneficial for the reconstruction of quantum states under imperfect conditions. 
	The key contributions of our work are summarized as follows:
	\begin{itemize}
		\item[$\bullet$]  We design a transformer-based autoencoder architecture, which learns to estimate an informative latent representation from imperfect measured data, benefiting the tomography of quantum states.
		\item[$\bullet$] We introduce a pre-training strategy to retrieve high-quality frequencies from imperfect data to relieve the ill-posedness and further enhance the expressiveness of the latent representation.
		\item[$\bullet$]  Numerical simulations demonstrate that ILR has significant potential in reconstructing density matrices over other NN-based methods. Experimental implementations on IBM quantum machines demonstrate the potential of ILR in compensating for noise in real quantum devices when directly using the model trained with simulated data.
	\end{itemize}
	
	\section{Related work}\label{Sect:related}
		In this section, we first review traditional QST methods and then introduce the NN-based QST method. Finally, we review transformer-based autoencoders.
		
	\subsection{Traditional QST} 
	Various approaches have been proposed for estimating quantum states based on measured statistics~\cite{qi2013quantum}. One commonly used method is the least-squares inversion that solves the inverse of linear equations which relates the measured quantities to the density matrix elements~\cite{opatrny1997least}. Maximum likelihood estimation (MLE) is another popular method that searches for the state estimate by maximizing the probability of observed data, but it involves a large number of nonlinear equations~\cite{jevzek2003quantum}. Linear regression estimation (LRE) transforms the problem into a linear model that can be efficiently solved but requires additional physical projection techniques to avoid non-physical quantum states~\cite{qi2013quantum,qi2017adaptive}. Bayesian tomography constructs a state using an integral averaging over all possible quantum states with proper weights~\cite{huszar2012adaptive}. While compressed sensing can estimate low-rank states  under an incomplete measurement setting~\cite{flammia2012quantum}, determining specific measurement operators restricts its applicability in real-world scenarios. When focusing on estimating partial properties, classical shadow estimation~\cite{huang2020predicting} and randomized benchmarking~\cite{magesan2012characterizing} are useful tools owing to their polynomial scale with system size.
	
	\subsection{NN-based QST}
	QST involves extracting useful information from measured statistics~\cite{jevzek2003quantum} and has been investigated using different neural networks. For instance, fully connected networks have been applied to approximate a function that maps measured statistics into quantum parameters~\cite{xu2018neural} and has been applied to denoise the state-preparation-and-measurement error~\cite{palmieri2020experimental}. A convolutional neural network has been introduced to reconstruct 2-qubit quantum states from an image of measured outcomes~\cite{lohani2020machine,lohani2021experimental}. Attention models have also been explored to capture long correlations among quantum data~\cite{cha2021attention,zhang2022transformer,zhong2022quantum}.
	While NNs have been applied to handle incomplete measurements~\cite{danaci2021machine}, the method relies on a separate statistical technique to infer missing values from existing data and the decoupling of various functionalities into different models. The capability of NNs has also been demonstrated by their application in reconstructing density matrices~\cite{ahmed2021quantum} and predicting statistics of new measurements~\cite{zhu2022flexible}. Compared to their methods utilizing a generative network, we design an enhanced NN architecture to learn an informative latent representation from imperfect measurement data, which benefits the reconstruction of quantum systems at different levels. Compared with classical shadow tomography which is specifically designed for predicting quantum properties~\cite{huang2020predicting}, our approach offers a unified framework that can be utilized to reconstruct density matrices and predict quantum properties. Compared to other works that utilize NNs for different tasks~\cite{carleo2017solving,torlai2018neural,schmale2022efficient}, our approach aims to learn an informative representation, which is shared between different tasks.

	\subsection{Transformer-based Autoencoder} 
	Autoencoding is a neural network technique to learn latent representations through an encoder that accepts raw input data and a decoder that reconstructs the input from the latent representation~\cite{hinton1993autoencoders}. It is a fundamental and powerful tool in deep learning with a wide range of applications, such as image denoising~\cite{kingma2013auto,ho2020denoising} and reconstruction~\cite{dong2015image}, natural language processing~\cite{cambria2014jumping}, and others. 
	The transformer architecture's self-attention mechanism~\cite{transformer2017} and pre-training technique, suitable for large dataset tasks, have popularized autoencoders in natural language processing (NLP)~\cite{devlin2018bert,raffel2020exploring, radford2018improving} and computer vision~\cite{alexey2020image,liu2021swin,he2022masked}, showing superiority to convolutional neural networks (CNNs)~\cite{lecun1998lenet-5,brown2020language}. Inspired by the potential of attention layers to capture long-range correlation among their constituent qubits~\cite{cha2021attention,zhong2022quantum}, we design a transformer-based autoencoder with a pre-training technique to reconstruct quantum states using the intermediate latent extracted from imperfect data.
	
	\section{Method}\label{Sec:method}
	
	In this section, we first introduce the basic concepts of QST and then present the ill-posed challenge for QST with imperfect data. Finally, a transformer-based autoencoder architecture with a pre-training strategy is proposed to learn the informative latent representation for QST. 
	
	\subsection{Preliminaries about QST}\label{sub:pre}  
	
	QST is a process of extracting useful information about a quantum state based on a set of measurements. For a $n$-qubit quantum system with dimension $d=2^n$, the full mathematical representation of a quantum state can be described as a density matrix $\rho\in \mathbb{C}^{d\times d}$. Measurement operators are usually positive-operator-valued measurements (POVMs) and can be represented as a set of positive semi-definite Hermitian matrices $\mathbb{O}_{i=1}^M \in \mathbb{C}^{d\times d}$ that sum to identity ($\sum_{i=1}^M\mathbb{O}_i=\mathbb{I}^{d\times d}$)~\cite{nielsen2010quantum}. According to Born's rule, when measuring the quantum state $\rho$ with the measurement operator $\mathbb{O}_i$, the true probability $p_i\in[0,1]$ of obtaining the outcome $i$ is calculated as 
	\begin{equation}
		p_i=\textup{Tr}(\mathbb{O}_i\rho).
		\label{eq:prob}
	\end{equation}
	In practice, a finite number of identical copies of a quantum state are used for measurements. In Fig.~\ref{fig:nnqst} (a), the statistical frequency for the $i$-outcome $f_i\in[0,1]$ is collected as
	\begin{equation}
		f_i=n_i / N,
		\label{eq:freq}
	\end{equation}
	where $N\in \mathbb{R}$ denotes the total number of copies of identical states and $n_i \in \mathbb{R}$ denotes the occurrences for the outcome $i$.
	As the true probability $p_i$ is directly inaccessible from experimental results, a natural solution is to leverage the measured frequency vector $\mathbf{f}=[f_1,f_2,...,f_M]^T\in\mathbb{R}^{M\times 1}$ which is a statistical approximation to the true probability vector $\mathbf{p}=[p_1,p_2,...,p_M]^T\in\mathbb{R}^{M\times 1}$, to infer underlying information about the quantum state $\rho$.  As illustrated in Fig.~\ref{fig:nnqst} (b), reconstructing quantum states can be solved by approximating a function that maps the frequency vector $\mathbf{f}$ into desired representations of a quantum state, e.g., a density matrix $\rho$. In addition, it is also useful to deduce significant properties such as purity, entanglement, and entropy~\cite{huang2020predicting}, which can be denoted as a real vector $\bm{\mu}=[\mu_1,\mu_2,...,\mu_k]^T\in \mathbb{R}^{k\times1}$, where $k$ denotes the number of properties.

	\subsection{The ill-posed QST problem}\label{sub:iqp}
	From Eq.~(\ref{eq:freq}), $f_i$ is a statistical approximation to the true probability $p_i$ and the accuracy of $f_i$ depends on the number of copies of identical states available to each measurement operator. To determine a unique state, the required number of linearly independent measurement operators scales exponentially with qubit number $n$~\cite{jevzek2003quantum}. In practical applications, the reconstruction of quantum states may be affected by imperfect conditions in two ways. i) Limited copies of identical states are assigned to each measurement operator among a complete set of measurements, and $\mathbf{f}=[f_1,f_2,...,f_M]^T\in \mathbb{R}^{M\times 1}$ tends to be under-sampled, resulting in a large gap between $\mathbf{f}$ and $\mathbf{p}$, hindering the estimation of quantum states~\cite{qi2017adaptive}. ii) Some measurements in a complete set may not be easily realized~\cite{liu2004quantum} in practical quantum systems, e.g., solid-state systems. In this case, an informationally incomplete set of measurements results in the incompleteness of $\mathbf{f}$. Suppose $m$ measurements are not accessible in the experiments, and then some elements are missing or masked among $\mathbf{f}$, ending up with a new vector $\mathbf{\widetilde{f}}\in \mathbb{R}^{(M-m)\times 1} $. The incompleteness of $\mathbf{\widetilde{f}}$ fails to characterize the Hilbert space of quantum systems and therefore leads to an inaccurate reconstruction of quantum states~\cite{danaci2021machine}.

	\begin{figure}
		\centering
		\includegraphics[width=0.45\textwidth]{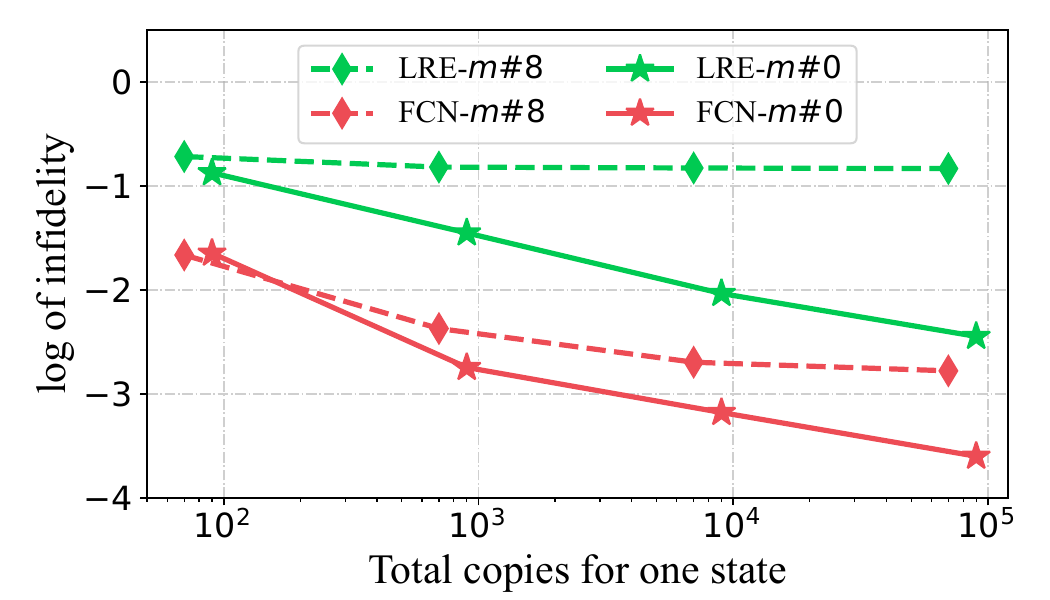}
		\caption{The ill-posed phenomena of QST. $m\#$ denotes the number of missed measurements. The point with the lower value of the log of infidelity is better. }
		\label{fig:ufp}
	\end{figure}

	The imperfect factors discussed above correspond to an ill-posed problem, which is a common issue in machine learning. In the context of quantum estimation tasks, ill-posedness results in an inaccurate or incomplete observation of quantum states, leading to errors in subsequent QST tasks. To provide a clear explanation, we conduct an initial experiment on 2-qubit pure states using the simplest fully connected network (FCN)~\cite{ma2021comparative} and a traditional linear regression estimation LRE~\cite{qi2013quantum}. The log of infidelity is measured as $\bar{F}=\log_{10}(1-F(\rho,\hat{\rho}))\in \mathbb{R}$, where  $F(\rho,\hat{\rho})=|\textup{Tr}(\sqrt{\sqrt{\rho}\hat{\rho}\sqrt{\rho}})|$ denotes the fidelity between the reconstructed state $\hat{\rho}$ and the original state $\rho$~\cite{nielsen2010quantum}. The numerical results in Fig.~\ref{fig:ufp} show a strong correlation between the infidelity and the copies of the identical states or the completeness of the measurement. Specifically, we observe a sharp decrease in infidelity as the number of copies increases, while masking $8$ operators among the original complete measurements (a total of 36 measurements) results in a large increase in infidelity. The underlying reasons might be that the noise addition and information loss in the imperfect measurement data result in a diminished representation of quantum states. The observed ill-posed challenge promotes the necessity of acquiring an informative latent representation that can potentially compensate for the imperfections in practical QST settings.
	
	\subsection{Learning an informative latent representation for QST with imperfect measurement data}\label{sub:ilr}
	
	\begin{figure*}[ht]
		\centering
		\includegraphics[width=0.9\textwidth]{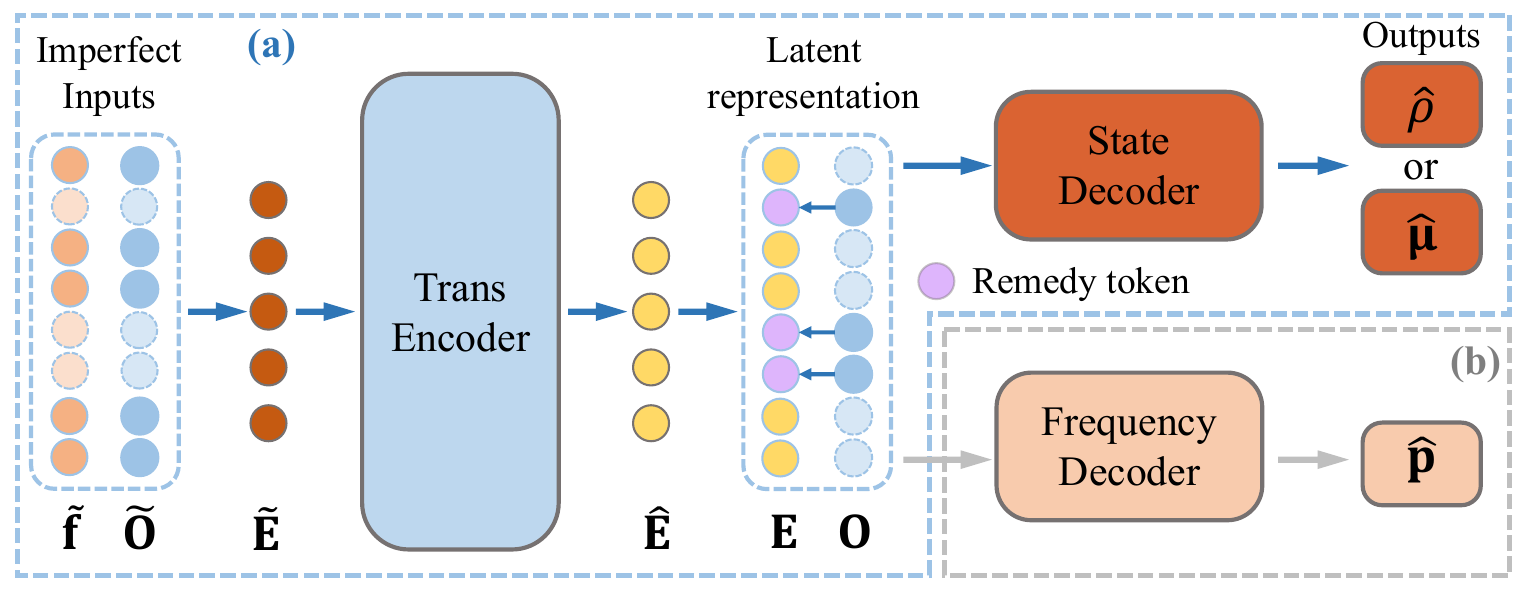}
		\caption{General schematic of learning an informative representation for quantum state tomography with imperfect measurement data. (a) The QST process aims to approximate a map from imperfect inputs (inaccurate and incomplete frequency vectors) to desired outputs (density matrices or quantum properties) through a transformer-based autoencoder and a state decoder. (b) The pre-training process aims to retrieve high-quality probabilities $\hat{\mathbf{p}}$ from masked frequency vector $\widetilde{\mathbf{f}}$ through a combination of a large encoder (share with (a)) and a small frequency decoder.
		}
		\label{fig:framework}
	\end{figure*}
	
	Given the representation reduction posed by ill-posed QST, it is imperative to design a robust neural architecture capable of learning a latent representation that might compensate for imperfect measurement data. Taking inspiration from the success of transformer-based autoencoders in learning latent representations from imperfect data in NLP and CV tasks, we propose an autoencoder to learn a latent representation from imperfect measurement data. The latent representation provides an informative feature that can be further leveraged to estimate quantum states by a decoder, including its density matrix and significant properties. 
	To further enhance the informative latent representation, a pre-training process is implemented to retrieve high-quality data from imperfect measurement data by augmenting a frequency decoder. The schematic of our approach is illustrated in Fig.~\ref{fig:framework}.
	
	\textbf{QST process using a transformer-based autoencoder. }
	In the QST process, a transformer-based autoencoder and a state decoder are combined to approximate a function that maps measurement frequencies to the desired parameters of quantum states, such as $\hat{\rho}$ (density matrix) or $\bm{\hat{\mu}}$ (physical properties). Here, we use a \textbf{hat} symbol $\hat{\cdot}$ to denote the reconstructed value from the neural networks. Hence, $\hat{\mathbf{p}}$, $\hat{\rho}$, and $\hat{\bm{\mu}}$ are utilized as the reconstructed outputs from the neural networks.
	In the encoding stage, we not only embed an imperfect frequency vector $\mathbf{\widetilde{f}}\in\mathbb{R}^{(M-m)\times 1}$ through a linear layer but also perform embedding on the corresponding incomplete operator matrix $\mathbf{\widetilde{O}} \in \mathbb{R}^{(M-m)\times 2d^2}$ (see \textbf{Remark} \ref{rm:operator}). Operator embedding leverages the capability of measurement operators~\cite{jevzek2003quantum,qi2013quantum} in capturing the correlations between different measurements, contributing to an enhanced latent representation of quantum states~\cite{cha2021attention,zhang2022transformer,zhong2022quantum}.
	Subsequently, these two embedded vectors are combined into a feature vector $\mathbf{\widetilde{E}}\in \mathbb{R}^{{(M-m)}\times L}$ (where $L$ denotes the embedding size). To reduce computational complexity by a factor of $d$, we reshape the feature vector $\mathbf{\widetilde{E}}$ into $\mathbb{R}^{G\times L}$, with $G=\frac{M-m}{d}$. Then $\mathbf{\widetilde{E}}$ is fed into a \emph{transformer-based encoder}, which consists of multiple stacked self-attention layers and feedforward layers. The self-attention mechanism allows the encoder to attend to different parts of the input feature vector and generate a new feature vector $\hat{\mathbf{E}}\in \mathbb{R}^{G\times L}$.

	In the decoding phase, we linearly embed the missed operators $(\mathbf{O} \setminus \mathbf{\widetilde{O}}) \in \mathbb{R}^{m\times 2d^2}$, which are omitted in the encoder input, as remedy tokens added to $\hat{\mathbf{E}}$. This results in a more informative latent representation $\mathbf{E} \in \mathbb{R}^{{(M/d)}\times L}$, corresponding to a complete frequency vector $\mathbf{f}\in\mathbb{R}^{M\times 1}$. Finally, we inject $\mathbf{E} \in \mathbb{R}^{{(M/d)}\times L}$ into a \emph{transformer-based state decoder} to deduce parameters of quantum states, including a density matrix $\hat{\rho}$ or a property vector $\bm{\hat{\mu}}$ that represents several physical features (see \textbf{Remark} \ref{rm:property}). 
	Taking into account the physical requirements of a density matrix, we introduce an additional post-processing module to generate a density matrix $\hat{\rho}$ from an intermediate real vector ($\bm\hat{\nu}$) (see \textbf{Remark} \ref{rm:density}). Finally, the parameters are trained using the mean squared error (MSE) loss between the prediction $\bm{\hat{\mu}}$ and the ground-truth $\bm\mu$, or $\bm{\hat{\nu}}$ and $\bm\nu$.
	
	\textbf{Pre-training process using a masked and asymmetrical autoencoder. }
	The transformer-based encoder utilized in the QST process is specifically designed to be expressive and proficient in extracting latent features from imperfect data. However, simply training the autoencoder model with raw data may not fully utilize the potential of the intermediate layers to abstract an informative latent representation from the imperfect measurement data. A high-dimensional ILR has the capability to capture a broader range of information pertaining to quantum states, resulting in enhanced accuracy during the reconstruction process (see Fig.~\ref{fig:ufp}). To obtain an improved ILR for QST with imperfect measurement data, we employ a pre-training strategy to guide the encoder to retrieve perfect measured data, i.e., true probabilities of a complete set of measurements. We reuse the transformer-based encoder from the QST process and combine it with a lightweight \emph{frequency decoder} to approximate the true probabilities from imperfect frequencies.
	
	Concretely, we draw the concept of a mask to deal with incomplete measurements and create a large number of imperfect training samples for pre-training. For example, suppose $m$ measurement operators are missing, a new frequency vector $\mathbf{\widetilde{f}}\in \mathbb{R}^{(M-m)\times 1} $ is masked out from the input frequency vector $\mathbf{f} \in \mathbb{R}^{M\times 1}$. Similarly, a corresponding mask is performed on the operator matrix $\mathbf{O}\in \mathbb{C}^{M\times 2d^2}$, ending up with a subset of the operator matrix $\widetilde{\mathbf{O}} \in \mathbb{C}^{(M-m)\times 2d^2}$. Here, the masking process can be sampled multiple times by randomly selecting $m$ from the integer set $\{0,1,2,3,..., M\}$, ending up with a large number of samples with imperfect input data. Following the transformer-based encoder reused from the QST process, we implement a specific \emph{transformer-based frequency decoder} to approximate the original probabilities $\mathbf{p}\in\mathbb{R}^{M\times 1}$ from masked frequencies ${\widetilde{\mathbf{f}}}\in \mathbb{R}^{(M-m)\times 1}$, rather than using the \emph{state decoder}. To optimize the parameters of the architecture, the pre-training process employs an MSE loss between the reconstructed probabilities $\hat{\mathbf{p}}$ and the true probabilities $\mathbf{p}$. Importantly, to prevent the expressiveness of the structure from transferring into the decoder and to ensure a sufficiently informative latent representation, the pre-training autoencoder is designed with an asymmetrical architecture, where the encoder is larger in scale compared to the decoder.

		\begin{remark}
		One operator $\mathbb{O}_i\in \mathbb{C}^{d\times d}$ can be rewritten as a real column vector $O_j \in \mathbb{R}^{2d^2\times 1}$ by splitting its real and imaginary parts. A set of operators $\{\mathbb{O}_1,\mathbb{O}_2,...,\mathbb{O}_M\}$ can be concatenated into a matrix $\mathbf{O}=[O_1,O_2,...,O_M] \in \mathbb{R}^{M\times 2d^2}$. 
		\label{rm:operator}
	\end{remark}
	
	\begin{remark}
		Generate a property vector $\bm{\mu}$: Let the number of properties be $K$, and then multiple properties form a real vector $\bm{\mu} \in \mathbb{R}^{K \times 1}$, with the ground-truth for each element calculated from the density matrix $\rho$ (Detailed information provided in \textbf{Appendix}~\ref{app:quantum}). 
		\label{rm:property}
	\end{remark}
	
	\begin{remark}
		A density matrix $\rho$ can be generated from a lower triangular matrix $\textstyle\rho_L\in \mathbb{C}^{d\times d}$ according to $\rho = \rho_L\rho_L^{\dagger}/ {\textup{Tr}(\rho_L\rho_L^{\dagger})}$. The matrices satisfy three requirements: (i) $\rho=\rho^{\dagger}$, (ii) $\textup{Tr}(\rho)=1$, and (iii) $\rho \geq 0$. Conversely, any $\rho$ can be decomposed using the Cholesky decomposition~\cite{higham1990analysis}, i.e., $\rho_L\rho_L^{\dagger} = \rho$. Furthermore, $\rho_L$ can be transformed into a real vector $\bm{\nu}\in \mathbb{R}^{d^2\times 1}$ by concatenating its real and imaginary parts. Hence, retrieving a real vector $\bm{\nu}$ is equivalent to recovering the density matrix $\rho$. 
		\label{rm:density}
	\end{remark}
	
	\subsection{Algorithm description}
	
	The pre-training process is first implemented to obtain a highly representative representation of quantum states. Then, complete training of the QST process is achieved by keeping the parameters of the transformer-encoder fixed and fine-tuning the parameters of the state decoder. This structure enables the preservation of the information that is learned from the pre-trained model, and the latent representation enables the state decoder to reconstruct quantum states with improved performance. 
	To simulate different incomplete measurement scenarios, we introduce two ways for pre-training: i) \emph{{Separate}}, fixed mask values are utilized for all samples, and ii) \emph{{Unified}}, where mask values are randomly selected for each training batch. For both cases, the same number of samples are used for training, while the learned representation regarding MSE can be different for training, which can be demonstrated in Fig.~\ref{fig:ab_main} of Section~\ref{sec:exp2}. Similarly, the QST process has the same two training options. The algorithm descriptions for the pre-training process and the QST process are presented in Algorithm \ref{alg1} and Algorithm \ref{alg2} (in Appendix G), respectively. The difference in this latent representation also influences the individual QST tasks; see detail in Fig.~\ref{fig:ab_main}.
	
	Now, there are four different strategies for training the whole model, including \textbf{S2S} (separate pre-training with separate tomography), \textbf{S2U} (separate pre-training with unified tomography), \textbf{U2S} (unified pre-training with separate tomography), and \textbf{U2U} (unified pre-training with unified tomography). We conduct a comprehensive comparison of 2-qubit pure states and determine \textbf{U2S} as a good strategy that balances accuracy and efficiency. 
	
	\begin{figure}[h]
		
			\begin{algorithm}[H]
				\small
				\caption{\small Description for pre-training.}\label{alg1}
				\begin{algorithmic}[1]
					\REQUIRE Operators $\mathbf{O}$, Frequencies $\mathbf{f}$, Ground-truth probabilities $\mathbf{p}$, Masked number list $\mathbf{m}$, Training epochs $T$, Dataset steps $S$. 
					\ENSURE Obtaining the pre-training model $\Phi({\Theta_{p}})$.
					\STATE Build pre-training model $\Phi({\Theta_{p}})$ with Encoder $\Phi({\Theta_{e}})$ and Frequency Decoder $\Phi({\Theta_{fd}})$;
					\STATE Initialize $\Phi({\Theta_{p}})$ with default Kaiming initialization;
					\STATE \textbf{for} $t=1,2,\cdots,T$ \textbf{do}
					\STATE \quad \textbf{for} $s=1,2,\cdots,S$ \textbf{do}
					\STATE \qquad \textbf{if} Training is \textit{\color{blue}{Separate}} \textbf{then}
					\STATE \quad\qquad Use the fixed mask number $m$;
					\STATE \qquad \textbf{if} Training is \textit{\color{blue}{Unified}} \textbf{then}
					\STATE \quad\qquad Random select $m$ from $\mathbf{m}$;
					\STATE \qquad Forward using $\mathbf{\widetilde{f}}$ and $\mathbf{\widetilde{O}}$ obtained with $m$;
					\STATE \qquad Backward by minimizing $MSE(\mathbf{\hat{p}}, \mathbf{p})$ in the \\\qquad pre-training process;
					\STATE \qquad Update ${\Theta_{e}}$ and ${\Theta_{fd}}$ with gradient descent; 
					\RETURN optimal $\Phi({\Theta_{p}})$.
				\end{algorithmic}
			\end{algorithm}
	\end{figure}

	\section{Experiments}\label{sec:results}
	
	In this section, we first provide the implementation details of ILR-QST. Then, we present the results of reconstructing density matrices as well as predicting quantum properties. 
	
	\subsection{Implementation details}\label{sub:setting}	
	\textbf{Quantum states.}
	In this work, we consider both pure states and mixed states. To generate pure random states, we utilize the Haar measure~\cite{mezzadri2006generate} to generate random unitary matrices. Let $\mathbb{U}^d$ be the set of all $d$-dimensional unitary operators, i.e., any $U \in \mathbb{U}^d$ satisfies $UU^{\dagger} =U^{\dagger}U = \mathbb{I}$. Owing to its invariance under group multiplication (i.e., any region of $\mathcal{U}^d$ carries the same weight in a group average)~\cite{mezzadri2006generate}, the Haar metric is utilized as a probability measure on a compact group. This property enables us to generate random unitary transformations that are further utilized to generate random pure states as $\rho_{haar} = U |\psi_0\rangle \langle \psi_0| U^{\dagger}$,
	where $|\psi_0\rangle$ is a fixed pure state. For mixed states, we consider the random matrix from the Ginibre ensembles~\cite{forrester2007eigenvalue} given by $\rho_G=\mathcal{N}(0,1)_d+\textup{i}\mathcal{N}(0,1)_d$,
	where $\mathcal{N}(0,1)_d$ represents random normal distributions of size $d \times d$ with zero mean and unity variance. Random density matrices using the Hilbert-Schmidt metric are given by $
		\rho_{ginibre}=\frac{\rho_G \rho_G^{\dagger}}{\textup{Tr}(\rho_G \rho_G^{\dagger})}$~\cite{ozawa2000entanglement}.
	
   \textbf{Quantum measurements.} For quantum measurements, we consider tensor products of Pauli matrices, which is also called cube measurement~\cite{de2008choice}. Denote the Pauli matrices as $\sigma=(\sigma_x,\sigma_y,\sigma_z)$, with
   $$\sigma_x=\left[\begin{array}{cc}
   			1 & 0\\
   			0 & -1
   		\end{array}\right],
   		\sigma_y=\left[\begin{array}{cc}
   			0 & -\rm{i}\\
   			\rm{i} & 0
   		\end{array}\right],
   		\sigma_z=\left[\begin{array}{cc}
   			0 & 1\\
   			1 & 0
   		\end{array}\right].$$
   	Let $|H\rangle$ and $|V\rangle$ be eigen-vectors of  $\sigma_z$, $|L\rangle$ and $|R\rangle$  be eigen-vectors of $\sigma_y$, and $|D\rangle$ and $|A\rangle$ be eigen-vectors of  $\sigma_x$. The POVM elements for one-qubit Pauli measurement are finally given as $P_{pauli}=\{|H\rangle \langle H|,|V\rangle \langle V|,|L\rangle \langle L|,|R\rangle \langle R|,|A\rangle \langle A|,|D\rangle \langle D|\}$. Generalizing the 1-qubit Pauli measurement  to  $n$-qubit using the tensor product, we obtain the following POVMs $
   		P= P_1 \otimes P_2\cdots  \otimes P_n, \quad \text{with} \quad P_i \in P_{pauli}$.
   	For $n$-qubits, there are $6^n$ POVMs involved in the cube measurement. Typically, the $6^n$ measurement operators can be arranged into several groups, with each group containing $d=2^n$ elements~\cite{adamson2010improving}. For example,  $|HH\rangle \langle HH|$,$|HV\rangle \langle HV|$,$|VH\rangle \langle VH|$, and $|VV\rangle \langle VV|$ can be allocated into one group. These four POVM elements compose a 2-qubit detector that can be experimentally realized on quantum devices. 
	
	\textbf{Measured data collection. }
	100,000 quantum states are randomly generated for both pure and mixed states, with 95,000 samples used for training and 5,000 for testing. In this work, training and testing data are drawn from the same distribution. We opt for a split with less testing data to save time on extensive comparisons of different methods. To address potential concerns about overfitting, we conduct additional experiments with various train/test splits and provide convergence curves using a 19:1 train/validation split, with detailed results included in Appendix~\ref{app:ratio}. In the simulation setting, one first obtains true probabilities according to Eq.~(\ref{eq:prob}) and simulates the observation process on desktop CPUs using multinomial distribution functions with true probabilities and total copies. Upon collecting the concurrence, the frequencies are obtained according to Eq.~(\ref{eq:freq}). 

	
	\textbf{Neural network architecture. } The model complexity of the transformer can be influenced by the embedding size, the multi-head number of attention modules, the expansion rate of FCN, and the number of stacked layers. Based on 2-qubit states, we compare the performance of different models (i.e., different numbers of parameters) and determine an intermediate model that achieves a good performance (please refer to Appendix~\ref{app:ratio} for detailed information):  1) Encoder: The transformer encoder consists of 8 stacked layers with 16 attention heads (each with a dimension of 32) and 256 hidden units. The input to the encoder is a sequence of the feature latent $\Tilde{E}_i$. The encoder outputs a sequence of latent features with the same dimension as the input. 2) Decoder: The parameters of the transformer model in the frequency encoder and state encoder are the same as those in the encoder, except for the number of layers. The frequency decoder employs one layer of the transformer block, whereas the state decoder consists of four transformer layers for reconstructing $\bm{\nu}$ (or equivalently $\rho$) and one transformer layer for predicting $\bm{\mu}$. Note that our methodology is versatile and can be employed for various numbers of qubits or different POVMs while maintaining the same architecture. The dimensions of the input and output are adapted based on the specific problem requirements. Additionally, certain hyperparameters, such as the number of layers or hidden units, may be tuned to optimize performance.
	
	\textbf{Training settings. }
	The total copies are allocated into several groups, with each containing $d$ measurement operators. We define the number of copies involved in $d$ operators as $N_t$ to denote an average value. 
	Hereafter, we use $m$ and $N_t$ to represent two imperfect factors in practical QST applications. Given a fixed $N_t$ (e.g., $N_t=100$), the pre-train process is performed on a large number of imperfect samples, with $m$ randomly sampled from its predefined set, e.g., $\{0,4,8,...\}$, while the QST process is then fine-tuned on a specific $m$ separately. In the training process, we utilize the default Adam optimizer, a batch size of 256, and an initial learning rate of $5E-03$. To improve convergence and reduce the risk of overfitting, we employ a strategy of cosine learning rate decay. The model is trained for 500 epochs with a warm-up strategy for the first 20 epochs, during which the learning rate is gradually increased from 0 to $5E-03$. 
	
		\begin{table*}
		\caption{The $MSE(\hat{\mathbf{p}}, \mathbf{p})$ in the pre-training process on 2-qubit pure states. \textbf{S}: Separate; \textbf{U}: Unified.}
		\label{tab:mse}
		\begin{center}
			\setlength{\tabcolsep}{1.0mm}{
					\begin{tabular}{c|cccc|cccc|cccc}
						\toprule[1pt]
						& \multicolumn{4}{c|}{$N_t=10$}          & \multicolumn{4}{c|}{$N_t=100$}         & \multicolumn{4}{c}{$N_t=1000$}        \\
						\midrule[1pt]
						$m$ & 0       & 8       & 12      & 16      & 0       & 8       & 12      & 16      & 0       & 8       & 12      & 16      \\
						\textbf{S} & 3.86E-4 & 8.37E-4 & 1.94E-03 & 4.59E-03 & 3.41E-6 & 2.24E-5 & 5.49E-4 & 2.31E-03 & 5.04E-7 & 1.23E-5 & 5.31E-4 & 2.26E-03\\
						\textbf{U} & 3.67E-4 & 9.58E-4 & 2.31E-03 & 4.91E-03 & 9.28E-6 & 4.59E-5 & 7.05E-4 & 2.52E-03 & 1.15E-6 & 1.31E-5 & 6.47E-4 & 2.46E-03\\
						\bottomrule[1pt]
			\end{tabular}}
		\end{center}
	\end{table*}
	
	\subsection{Training strategies of pre-training and QST}\label{sec:exp2}
	
	\begin{figure*}
		\centering
		\begin{minipage}[b]{0.32\textwidth}
			\centering
			\includegraphics[width=0.9\textwidth]{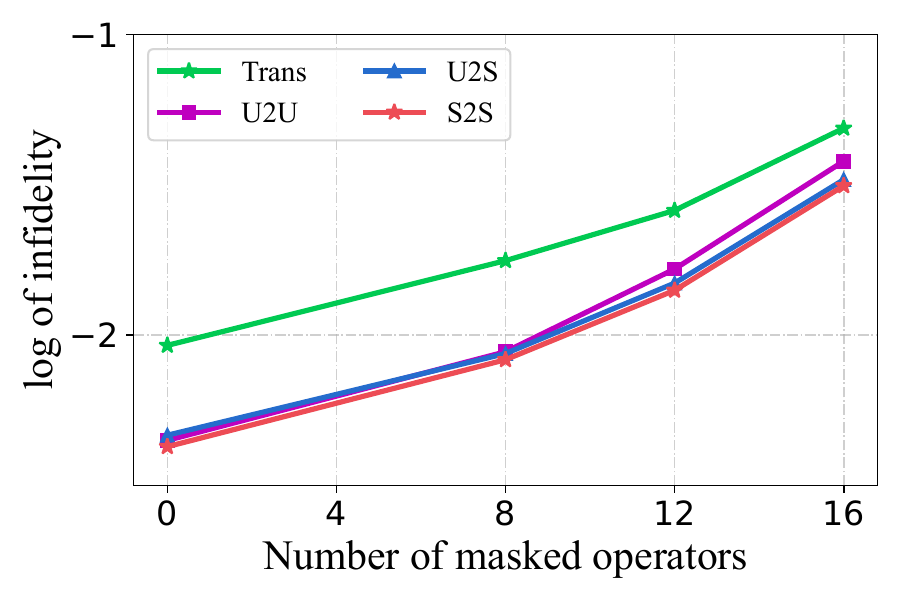}
			\centerline{(a) $N_t=10$}
			\label{sfig:ab1}
		\end{minipage}
		\hfill
		\begin{minipage}[b]{0.32\textwidth}
			\centering
			\includegraphics[width=0.9\textwidth]{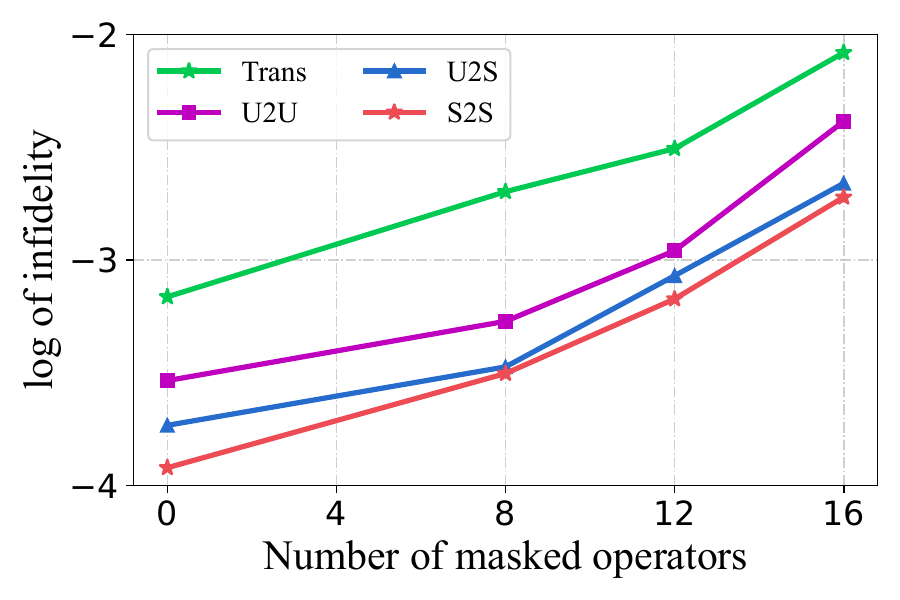}
			\centerline{(b) $N_t=100$}
			\label{sfig:ab2}
		\end{minipage}
		\hfill
		\begin{minipage}[b]{0.32\textwidth}
			\centering
			\includegraphics[width=0.9\textwidth]{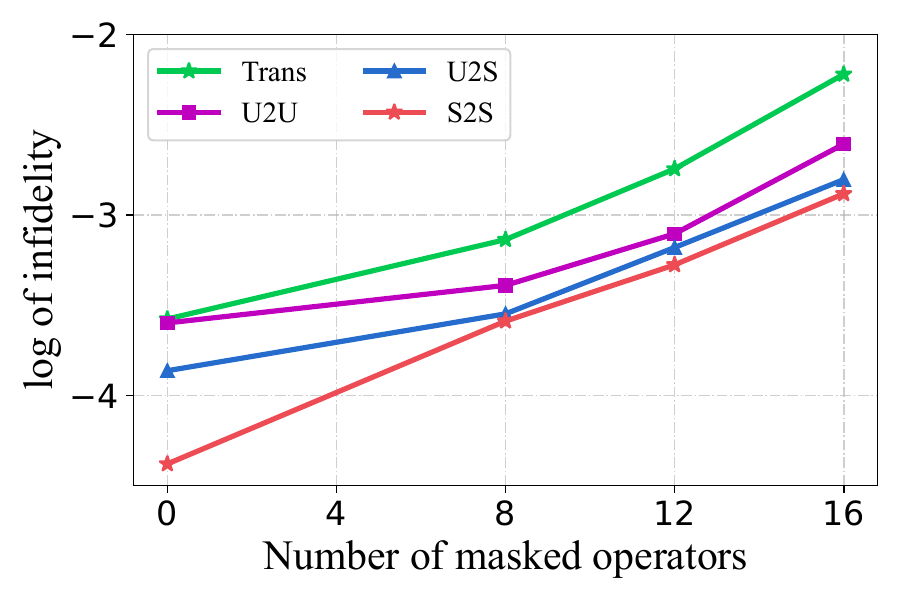}
			\centerline{(c) $N_t=1000$}
			\label{sfig:ab3}
		\end{minipage}
		\caption{Comparison of different training strategies on 2-qubit pure states.}
		\label{fig:ab_main}	
	\end{figure*}

	To evaluate the expressiveness of ILR learned during the pre-training process, we implement a comparison of two strategies ways on 2-qubit pure states,  with results provided in Table \ref{tab:mse}. Generally, the \emph{unified} strategy in the pre-training process demonstrates comparable performance to the \emph{separate} strategy in most cases. Although the \emph{separate} strategy shows superiority when $m=0$, it comes at the expense of higher computational complexity. Hence, the \emph{unified} strategy achieves a balance between computational cost and learning effectiveness. 

	Next, we explore the training strategies when combining pre-training and QST. Theoretical considerations suggest four possible strategies, but we exclude the \textbf{S2U} due to its impracticality and meaninglessness. We consider the remaining three strategies with the sole transformer model (marked as Trans in the following figures) and investigate their performance on 2-qubit pure states. The numerical results are summarized in Fig.~\ref{fig:ab_main}, where \textbf{S2S} achieves the best reconstruction accuracy among the implemented three strategies, as indicated by the lowest log of infidelity. There is no significant difference between the three strategies when copies of identical states are very limited, e.g., $N_t=10$. However, \textbf{S2S} exhibits clear superiority over other methods when $N_t=1000$. When considering a practical situation with limited copies, for example, $N_t=100$, \textbf{U2S} achieves a comparative performance to \textbf{S2S} with a slight disadvantage. Hence, \textbf{U2S} provides a useful strategy that balances accuracy and efficiency and we utilize it for the following experiments.
	
	\subsection{Reconstructing density matrices}\label{sub:main}
	
	\begin{figure*}[h]
	\begin{minipage}[b]{0.48\textwidth}
		\centering
			\includegraphics[width=0.98\textwidth]{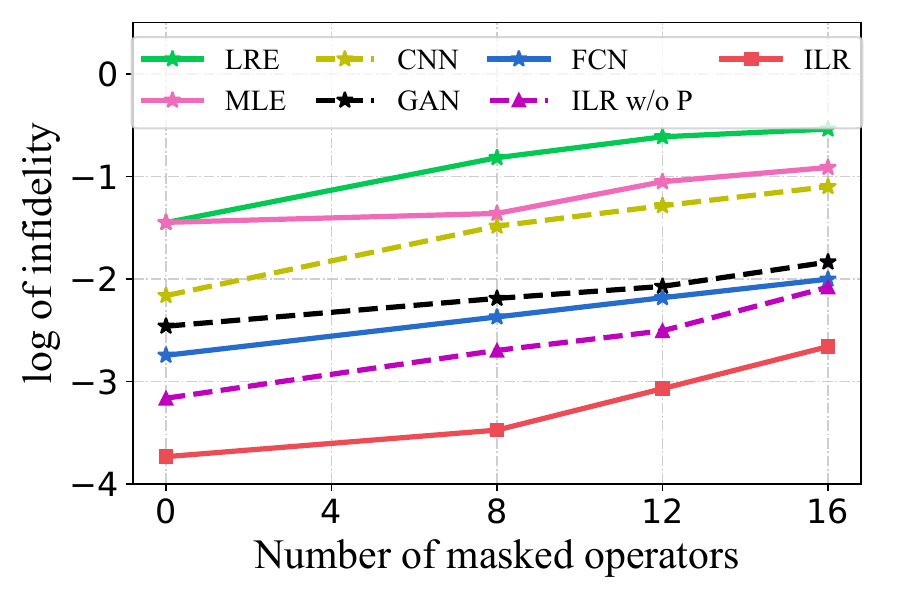}
				\centerline{(a) 2-qubit states}
	\end{minipage}
    \hfill 
     \begin{minipage}[b]{0.48\textwidth}
		\includegraphics[width=0.98\textwidth]{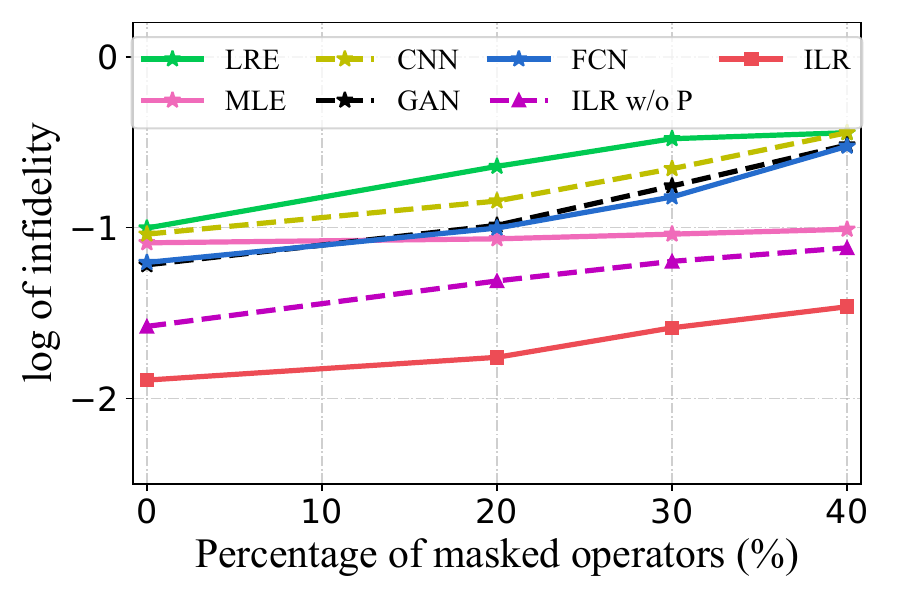}
			\centerline{(b) 4-qubit states}
	\end{minipage}
	\caption{Comparison results of reconstructing density matrices for pure states ($N_t=100$) with other methods including LRE, MLE, CNN, GAN, and FCN. ``ILR w/o P" indicates ILR without the pre-training strategy.}
	\label{fig:purestates}	
\end{figure*}

\begin{table*}[h]
	\centering
	\caption{\textcolor{black}{Comparison of parameters and inference time for ILR and other methods used in Fig.~\ref{fig:purestates}. Infer-CPU/GPU denotes the inference process performed on CPU/GPU. Despite the need for training and parameters in NN approaches, they consistently demonstrate superior performance to traditional methods. Notably, our ILR model outperformed FCN approaches in terms of performance with comparable parameters and inference time.}}
	\setlength{\tabcolsep}{1.0mm}{
			\begin{tabular}{c|ccccc|ccccc}
				\toprule[1pt]
				& \multicolumn{5}{c|}{2 qubit} & \multicolumn{4}{c}{4 qubit}\\
				\midrule[1pt]
				Methods & Infidelity & Params  & Train-GPU & Infer-CPU& Infer-GPU& Infidelity & Params & Train-GPU & Infer-CPU & Infer-GPU\\
				LRE & 3.54E-2 & N/A & N/A & 4.57ms & N/A & 9.93E-2& N/A &  N/A & 671ms& N/A \\
				MLE & 3.56E-2 & N/A & N/A & 86.6ms & N/A & 8.14E-2& N/A  & N/A & 793ms & N/A\\
				\midrule[1pt]
				FCN & 1.80E-03  & 211K& 1.0h & 5.42ms & 1.30ms & 6.25E-2 & 595K & 1.5h & 20.35ms & 1.88ms \\
				ILR  & 1.85E-4  & 218K & 2.5h & 6.92ms & 1.38ms & 1.28E-2  & 257K & 6.0h & 23.71ms  & 2.23ms\\
				\bottomrule[1pt]
	\end{tabular}}
	\label{table:paracompare}
\end{table*}

\begin{table*}[h]
	\centering
	\caption{Statistical fidelities (from 50 samples) of ILR (model trained in Fig.~\ref{fig:purestates} (a)) and other methods on IBM quantum computer. Here a higher value of fidelity means a better performance. }
	\begin{tabular}{cccccc}
		\toprule[1pt]
		& \multicolumn{2}{c}{Simulation data}             & \multicolumn{3}{c}{Experimental data (\emph{ibmq\_{manila}})}             \\
		\midrule[1pt]
		& ILR      & LRE      & ILR  &  LRE  & IBM-built in  \\
		\midrule[1pt]
		Min        & 0.999463 & 0.911296 & 0.930446 & 0.833402 & 0.819149 \\
		Max        & 0.999982 & 0.995680 & 0.993310 & 0.960062 &  0.957353 \\
		Mean    & 0.999828 & 0.972375 & 0.973660 & 0.901905 &  0.892300 \\
		Variance   & 2.42E-08 & 6.01E-04 & 3.16E-04 & 1.06E-03 &  1.15E-03 \\
		\toprule[1pt]
	\end{tabular}
	\label{tabel:experiment}
\end{table*}

\begin{figure}
	\centering	
	\begin{minipage}[b]{0.45\textwidth}
		\centering
		\includegraphics[width=0.95\linewidth]{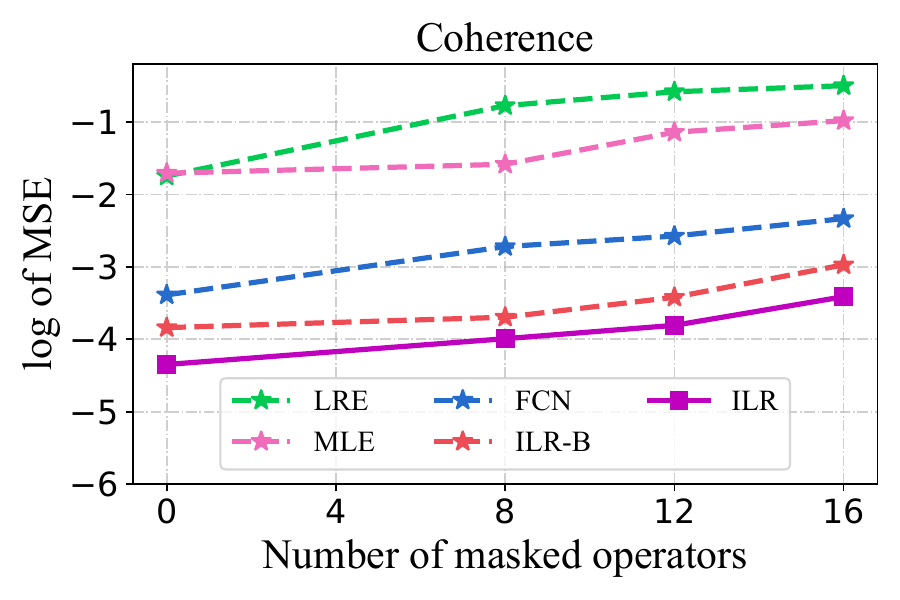}\label{sfig:pp11}
	\end{minipage}
	\vfill
	\begin{minipage}[b]{0.45\textwidth}
		\centering
		\includegraphics[width=0.95\linewidth]{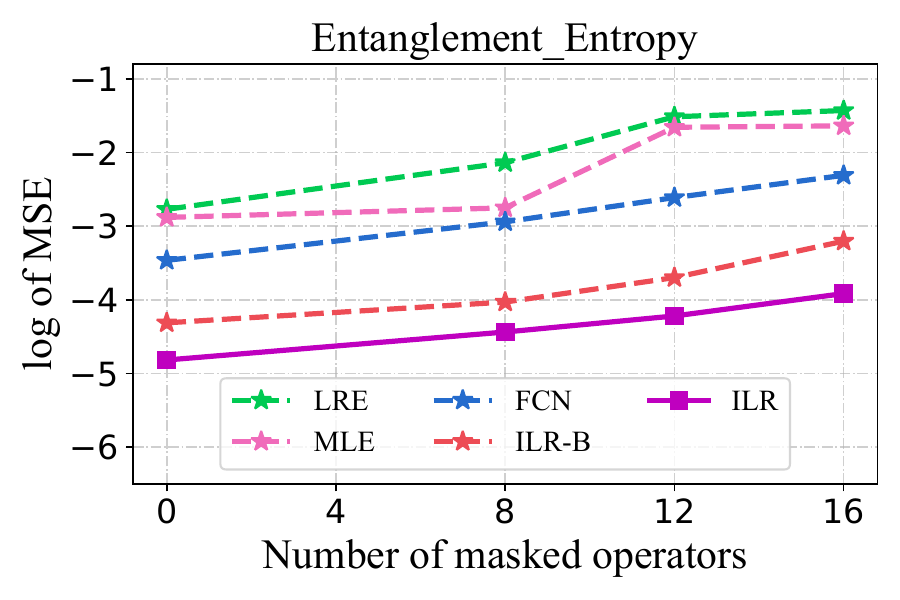}\label{sfig:pp12}
	\end{minipage}
	\vfill
	\begin{minipage}[b]{0.45\textwidth}
		\centering
		\includegraphics[width=0.95\linewidth]{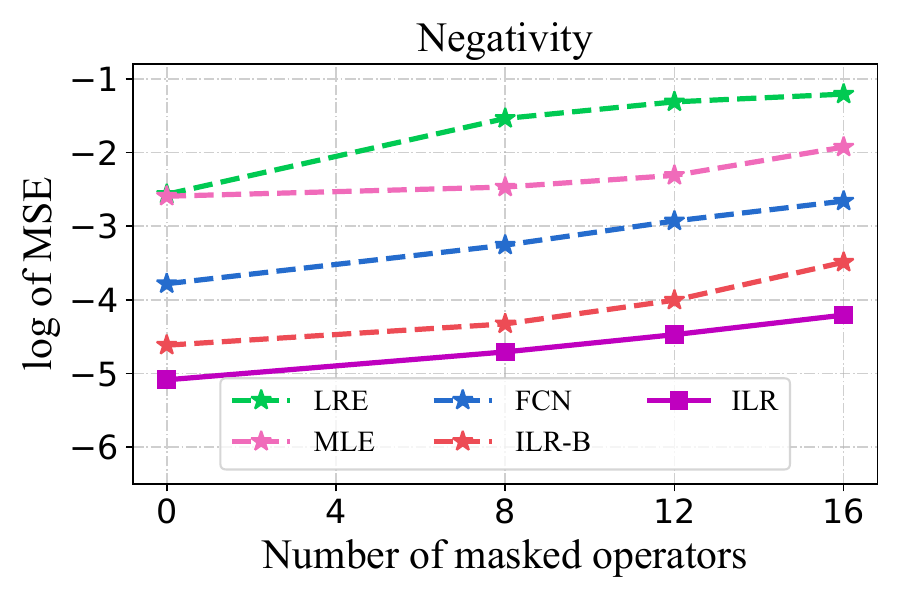}\label{sfig:pp13}
	\end{minipage}
	\caption{Results of predicting properties for 2-qubit pure states ($N_t=100$). In ILR-B, FCN, MLE, and LRE (dash lines), quantum properties are calculated from the reconstructed density matrices.}
	\label{fig:property-pure-2qubit}	  
\end{figure}

\begin{figure}
	\centering	
	\begin{minipage}[b]{0.45\textwidth}
		\centering
		\includegraphics[width=0.95\linewidth]{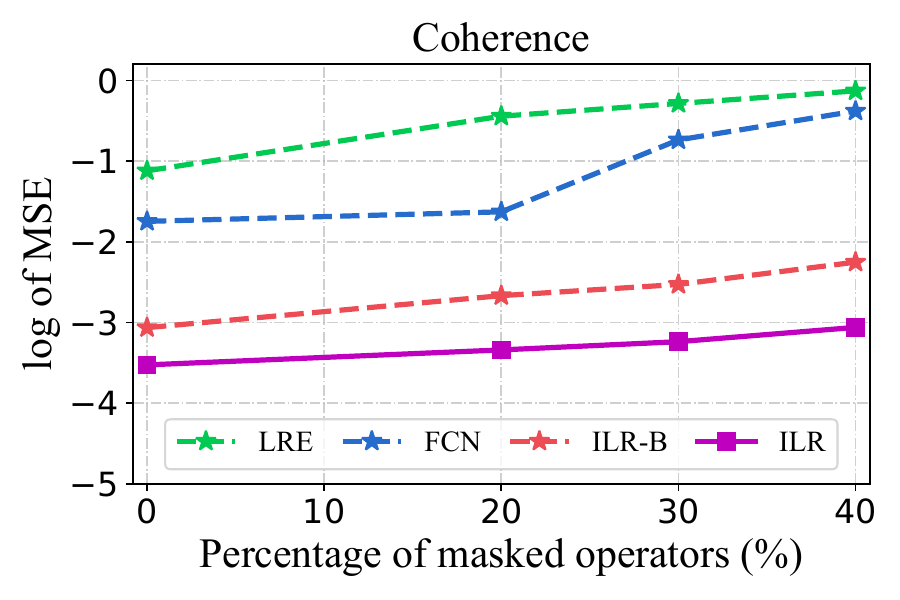}
	\end{minipage}
	\vfill
	\begin{minipage}[b]{0.45\textwidth}
		\centering
		\includegraphics[width=0.95\linewidth]{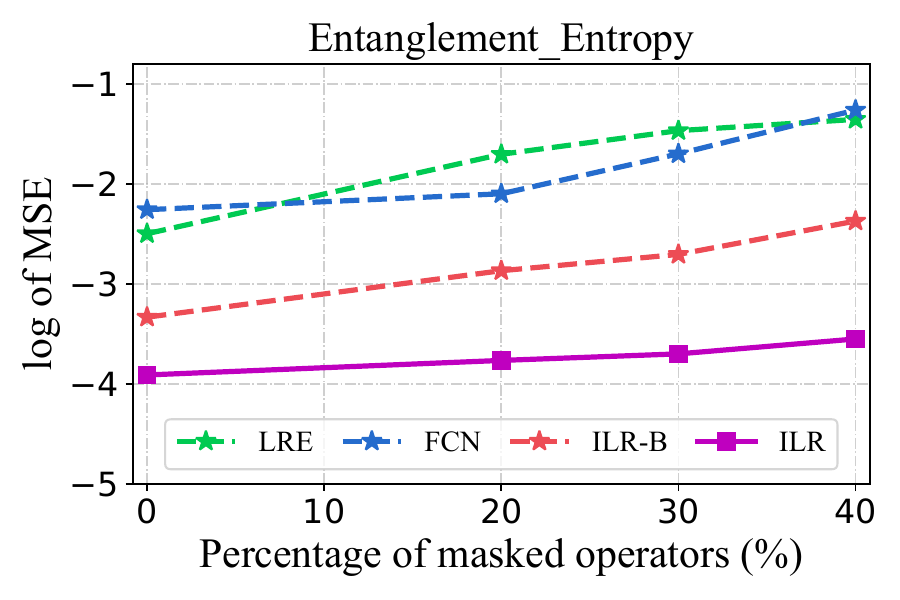}
	\end{minipage}
	\vfill
	\begin{minipage}[b]{0.45\textwidth}
		\centering
		\includegraphics[width=0.95\linewidth]{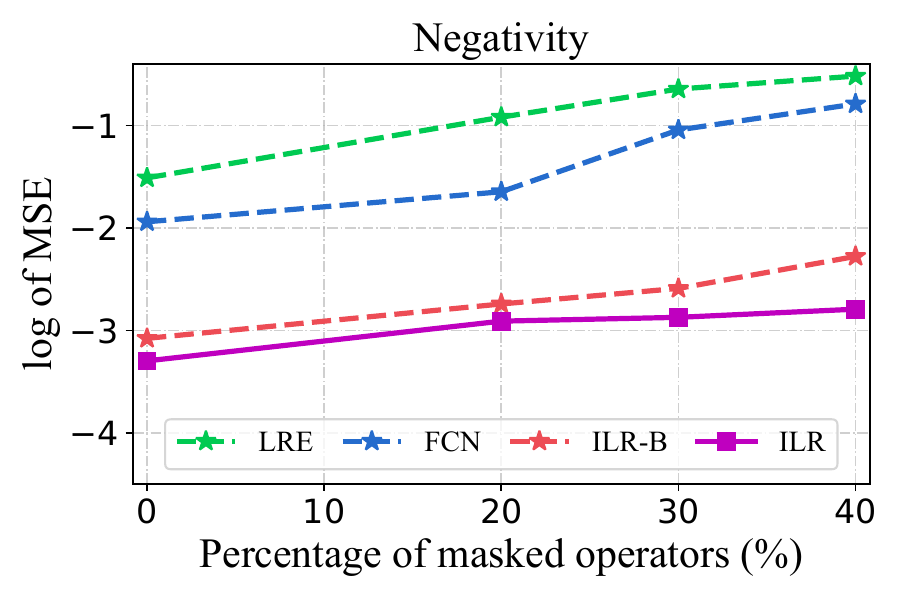}
	\end{minipage}
	\caption{Results of predicting properties for 4-qubit pure states ($N_t=100$). In ILR-B, FCN, MLE, and LRE (dash lines), quantum properties are calculated from the reconstructed density matrices.}
	\label{fig:property-pure-4qubit}	  
\end{figure}

Here, we implement the FCN~\cite{ma2021comparative}, LRE~\cite{qi2013quantum}, and MLE~\cite{jevzek2003quantum} and CNN-based QST~\cite{lohani2020machine} and GAN-based QST~\cite{ahmed2021quantum} for comparison. For NN-based methods, we take the same setting for the number of hidden layers and hidden neurons from the involved references. 
To conduct an overall comparison, we compare the number of parameters, and inference time (GPU/CPU) for the proposed ILR and other methods. The simulation platform consists of an 8-core Intel(R) Xeon(R) W-2145 CPU @ 3.70GHz, 64G DDR4 memory size, and Quadro RTX 4000 with 8G. For inference time (CPU/GPU), we conduct 5,000 times of reconstructing density matrix and report the average time per inference in Table~\ref{table:paracompare}. Owing to the connections between layers in NNs, the involved number of parameters is naturally large, which is common in reconstructing quantum states, including for 2 qubit states~\cite{danaci2021machine,lohani2021experimental,koutny2022neural}. In typical FCN-based methods, each neuron connects to every neuron in the previous layer, causing the number of parameters to increase significantly with the size of the connected layers but less computation. In contrast, the Transformer architecture, though inherently more complex with advanced blocks like attention, across different positions in the input sequence~\cite{transformer2017}. This leads to slightly longer computation times compared to FCNs, but with only 200K parameters, the Transformer remains relatively lightweight in the context of modern deep-learning models. Compared to other advanced models that often exceed millions of parameters, our model remains quite lightweight with a balance of complexity and efficiency.

Due to the exponential scaling issues in full QST, we implement simulations on low qubits in this work. The comparison results for 2-qubit and 4-qubit pure states with $N_t=100$ are summarized in Fig.~\ref{fig:purestates}, where the log of the infidelity of the four methods increases with the number of masked operators, which is in agreement with theoretical expectations. The proposed approach (ILR) exhibits superiority over the classical methods, i.e., LRE and MLE, and also outperforms other NN-based methods, i.e., FCN, CNN, and GAN-based QST methods. Results for mixed states and different $N_t$ are summarized in \textbf{Appendix}~\ref{app:density}. Those results suggest that the ILR is effective in reconstructing density matrices from imperfect measurement data. 

We also perform experiments on IBM quantum machine $\emph{ibmq\_{manila}}$ and results for 50 samples are provided in Table~\ref{tabel:experiment}. Note that our ILR model is trained using simulated data that consider noise from the measurement shots, i.e., the number of shots is the same in simulated and experimental settings. The results show the excellent robustness of our ILR model in dealing with experimental data, compared with LRE and the IBM-built method. However, one should note that without considering the errors from the state preparation and the measurement operator in real quantum devices~\cite{rambach2022efficient,ivanova2023optimal}, the reconstruction accuracy can be hindered, which motivates us to design a more powerful method to overcome such limitations~\cite{palmieri2020experimental}.

\subsection{Predicting quantum properties}\label{sub:pqp}

\begin{table*}[h]
	\centering
	\caption{MSE of predicting properties on 8/12-qubit GHZ and W states with $N_t=100$. Mask Level (1,2,3,4): (0, 16, 24, 32) out of 63 for 8-qubit; (0, 24, 36, 48) out of 99 for 12-qubit.} 
		\begin{tabular}{ccccc}
			\toprule[1pt]
			& \multicolumn{2}{c}{GHZ states (Coherence/Entanglement)} & \multicolumn{2}{c}{W states (Coherence/Entanglement)} \\
			\midrule[1pt]
			Mask Level  &  8-qubit   & 12-qubit     & 8-qubit    & 12-qubit \\
			\midrule[1pt]
			1          & 2.01E-03/8.02E-04  & 1.12E-02/2.41E-03       & 8.22E-03/3.02E-03      & 7.20E-02/2.86E-03 \\
			2             & 1.26E-02/3.18E-03   & 5.36E-02/3.13E-03       & 9.37E-03/3.18E-03     &7.19E-02/2.93E-03\\
			3          & 1.64E-02/3.23E-03     & 6.87E-02/3.11E-03      & 1.07E-02/3.23E-03     & 7.18E-02/2.93E-03 \\
			4          & 2.44E-02/3.37E-03   & 8.82E-02/4.06E-03       & 1.25E-02/1.93E-03  & 7.23E-02/2.87E-03  \\
			\bottomrule[1pt]
			\label{table:scalability-all}
	\end{tabular}
\end{table*}

\begin{figure*}
    \centering
    \includegraphics[width=0.9\linewidth]{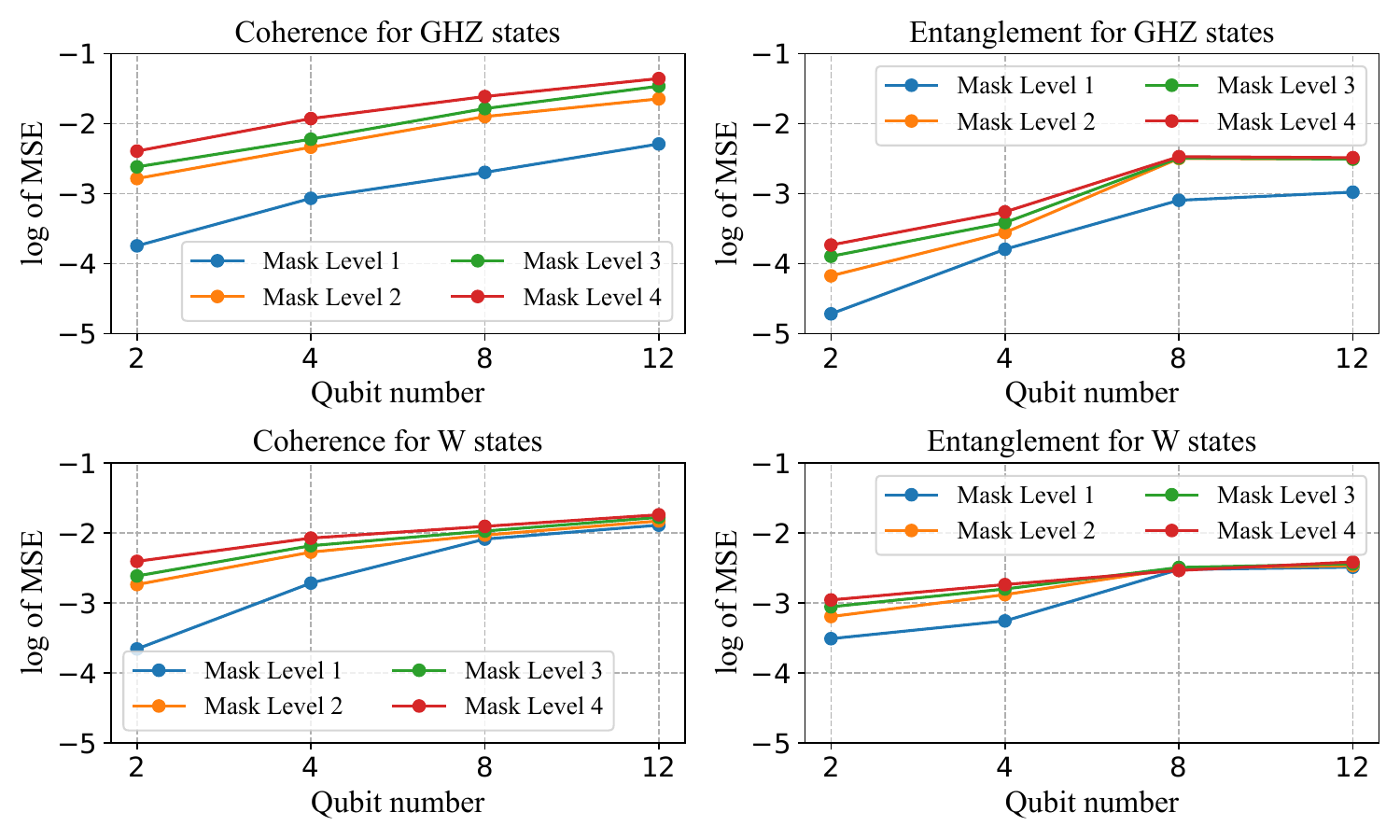}
    \caption{Numerical results of predicting properties for different numbers of qubits.}
    \label{fig:scaling}
\end{figure*}


Apart from directly using ILR to predict quantum properties, we introduce a baseline with properties calculated from density matrices using ILR in Subsection \ref{sub:main} (termed ILR-B). Please refer to~\textbf{Appendix}~\ref{app:quantum-2bit-nt1000} for their detailed comparsion. Then, we also summarize the results of calculating properties from reconstructed density matrices using FCN, LRE, and MLE. To distinguish them from ILR, dash lines are utilized for the above four methods. The results for 2-qubit and 4-qubit states at $N_t=100$ are provided in Fig.~\ref{fig:property-pure-2qubit} and Fig.~\ref{fig:property-pure-4qubit}, with results of 2-qubit mixed states presented in \textbf{Appendix}~\ref{app:predictingproperties}. The proposed ILR directly predicts properties with the highest accuracy, with great superiority over the indirect methods that first recover the density matrix using traditional algorithms and then calculate properties. Those results demonstrate the informative representation has great potential to deduce a partial description of quantum states under imperfect conditions.

To demonstrate the applicability of ILR to high-dimensional states, we consider physical states with specific patterns, i.e., locally rotated GHZ states of the form $\otimes_{i=1}^{n}U_i |\psi_{GHZ}\rangle$ with  $|\psi_{GHZ}\rangle=\frac{1}{\sqrt{2}}(|00...00\rangle+|11...11\rangle)$, where $U_i$ denotes a random unitary transformation in the $i$-th qubit. We also consider locally rotated W states of the form $\otimes_{i=1}^{n}U_i |\psi_{W}\rangle$ with  $|\psi_{W}\rangle=\frac{1}{\sqrt{n}}(|100...00\rangle+|010...00\rangle+...+|000...10\rangle+|000...01\rangle)$. We utilize two-qubit Pauli measurements on nearest-neighbor qubits, with $(n-1)*9*4$ types of measurement~\cite{zhu2022flexible}. Owing to the high computation and storage of high qubits, we utilize 10000/1000 states for 8/12 qubits, with a training and testing ratio of 9:1. From Table~\ref{table:scalability-all}, our method has good stability in predicting the properties of quantum states. To demonstrate the performance with increasing qubits, we present the numerical results of different qubits in Fig.~\ref{fig:scaling}, where the total samples are 10000 with a training/testing ratio being 9:1. This demonstrates that ILR has great potential to deduce a partial description of quantum states under imperfect conditions.



\section{Conclusion}\label{sec:conclusion}
In this paper, we investigated QST with imperfect measurement data, i.e., a few copies of identical states to approximate true probabilities or incomplete measurements that theoretically fail to specify a unique solution. Drawing the similarity of incompleteness and mask, we designed a masked autoencoder to map the raw measured frequencies into high-quality probabilities. By pre-training the masked autoencoder with a large number of samples obtained from ill-posed scenarios, a highly informative latent representation has been extracted and utilized to efficiently reconstruct quantum states from imperfect measured outcomes. The numerical simulations and experimental implementations demonstrate that ILR provides a robust and efficient solution for QST with imperfect measurement data. However, it should be pointed out that our method does not solve the exponential scaling challenge for full tomography of quantum states. Numerical results demonstrate the performance of partial estimation is robust across different qubits. Our future work will focus on dealing with real errors in quantum devices or neural network-based quantum process tomography. 
	
\appendix
\section{Appendices}

\subsection{Quantum properties}\label{app:quantum}

\begin{table}[h]
		\centering
		\vspace{-0.1cm}
		\caption{Descriptions for quantum properties.}
		\vspace{0.3cm}
		\renewcommand\arraystretch{1.2}
			\begin{tabular}{l|c|c} 
				\midrule[1pt]
				Name & Definition & Range \\
				\midrule[1pt]
				Purity &  $\textup{Tr}(\rho^2)$ & $[0,1]$\\
				\midrule[1pt]
				Entropy & $\textbf{S}(\rho):=-\textup{Tr}(\rho \ln(\rho))$ &$ [0,\ln(d)]$\\
				\midrule[1pt]
				Coherence & $S(\rho_{diag})-S(\rho)$ &$ [0,\ln(d)]$\\
				\midrule[1pt]
				Entanglement \\ entropy & $S(\textup{Tr}_A(\rho))$ &$ [0,\ln(d)/2]$ \\
				\midrule[1pt]
				Negativity & $\frac{||\rho^{\Gamma_A}||_1-1}{2}$ &$ [0,0.5]$\\
				\midrule[1pt]
				Concurrence & $\max(0,\lambda_1-\lambda_2-\lambda_3-\lambda_4)$ &$ [0,1]$\\
				\midrule[1pt]
			\end{tabular}
		\begin{tablenotes}
			\footnotesize
			\item Note: $\textup{Tr}_A(\rho)$ represents partial trace of $\rho$ over subsystem $A$,  $\rho^{\Gamma_A}$ is partial transpose with respect to subsystem $A$. $\rho_{diag}$ denotes the matrix by deleting all the off-diagonal elements. 
		\end{tablenotes}
		\label{table:property}
	\end{table}
	
	In this work, we consider six properties that are widely used in quantum information processing tasks. Pure states have a fixed purity of 1 and a fixed entropy of 0, which can be proven by simple calculation from the equations in Tabel \ref{table:property}. Coherence is commonly measured by the sum of squared values of non-diagonal elements among the density matrix, i.e., $\sum_{i,j, i \neq j} |\rho_{ij}|$, which can be extremely large for high-dimensional cases. Instead, we consider relative coherence regarding the entropy~\cite{baumgratz2014quantifying}. The von Neumann entanglement entropy is defined as the von Neumann entropy of either of its reduced states since they are of the same value, which can be proved from Schmidt's decomposition of the state with respect to the bipartition. Concurrence is originally defined for 2-qubit mixed states, with $\{\lambda_i\}$ being the eigenvalues (decreasing) of the Hermitian matrix $R=\sqrt{\sqrt{\rho}\rho_y\sqrt{\rho}}$ with $\rho_y=(\sigma_y\otimes \sigma_y) \rho^* (\sigma_y\otimes \sigma_y)$, with the complex conjugation $^*$ taken in the eigenbasis of the Pauli matrix $\sigma _{z}$. A generalized version of concurrence for multiparticle pure states in arbitrary dimension is defined as $\sqrt{2(1-\textup{Tr}(\textup{Tr}_A(\rho)^2))}$. Among the six properties in Tabel \ref{table:property}, the latter three properties are designed to evaluate the entanglement of quantum states. Specifically, their values achieve the lowest value of zero for separable states and achieve the highest value for maximally entangled states, e.g., Greenberger–Horne–Zeilinger states.  
	
%

	\subsection{Additional results for reconstructing density matrices}\label{app:density}
	
	\begin{figure*}
	\centering
	\begin{minipage}[b]{0.32\textwidth}
		\centering
		\includegraphics[width=0.9\textwidth]{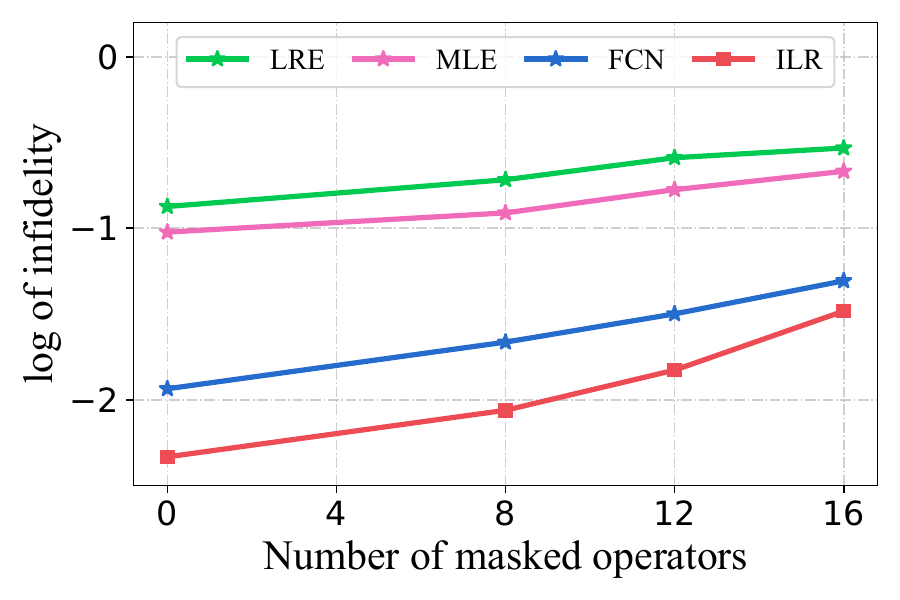}
		\centerline{(a) $N_t=10$}
		\label{pure-N10}
	\end{minipage}
	\hfill
	\begin{minipage}[b]{0.32\textwidth}
		\centering
		\includegraphics[width=0.9\textwidth]{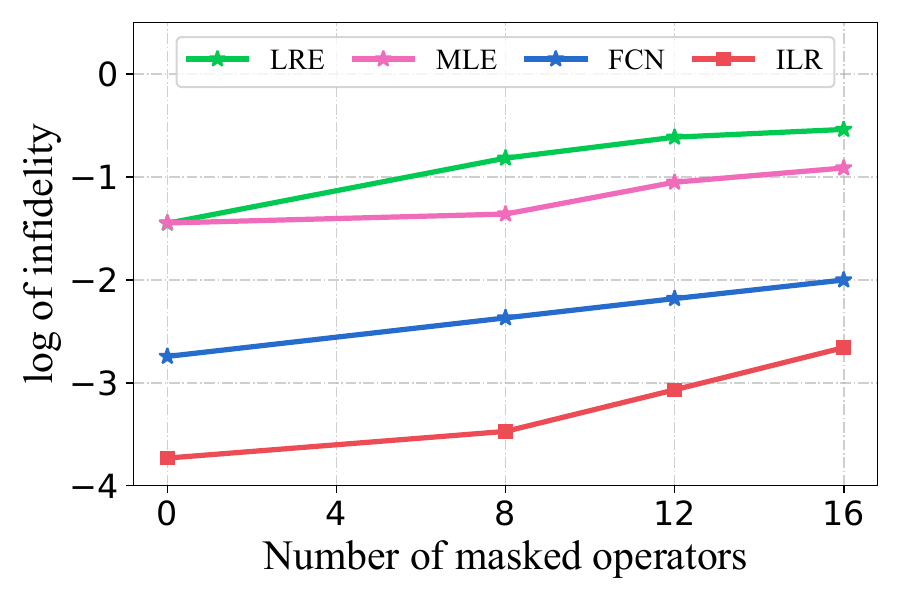}
		\centerline{ (b) $N_t=100$}
		\label{pure-N100}
	\end{minipage}
	\hfill
	\begin{minipage}[b]{0.32\textwidth}
		\centering
		\includegraphics[width=0.9\textwidth]{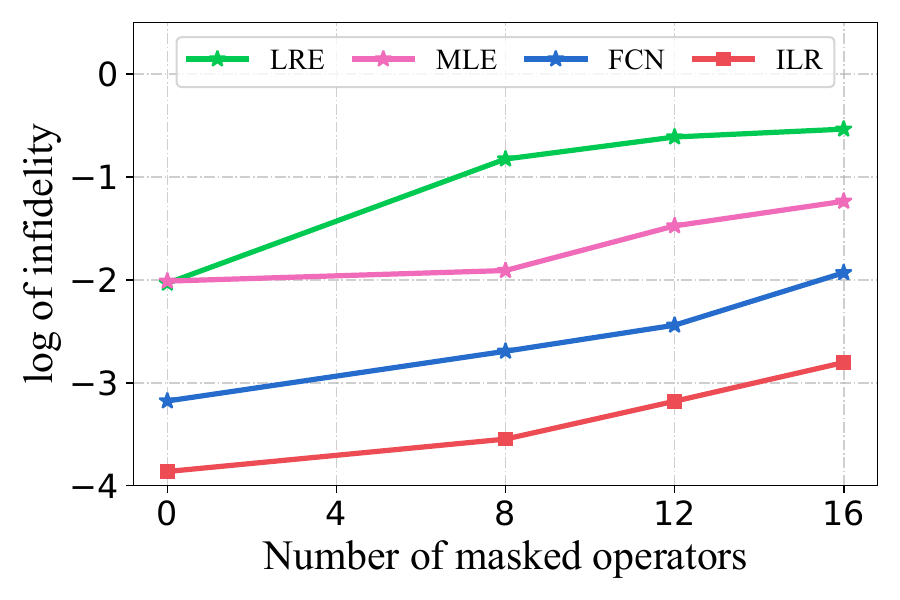}
		\centerline{ (b) $N_t=1000$}
		\label{pure-N1000}
	\end{minipage}
	\caption{Comparison results of different $N_t$ for 2-qubit pure states}
	\label{fig:pure_Nt}	
\end{figure*}

	\begin{figure*}[h]
		\centering	
		\begin{minipage}[b]{0.32\textwidth}
			\centering
			\includegraphics[width=0.9\textwidth]{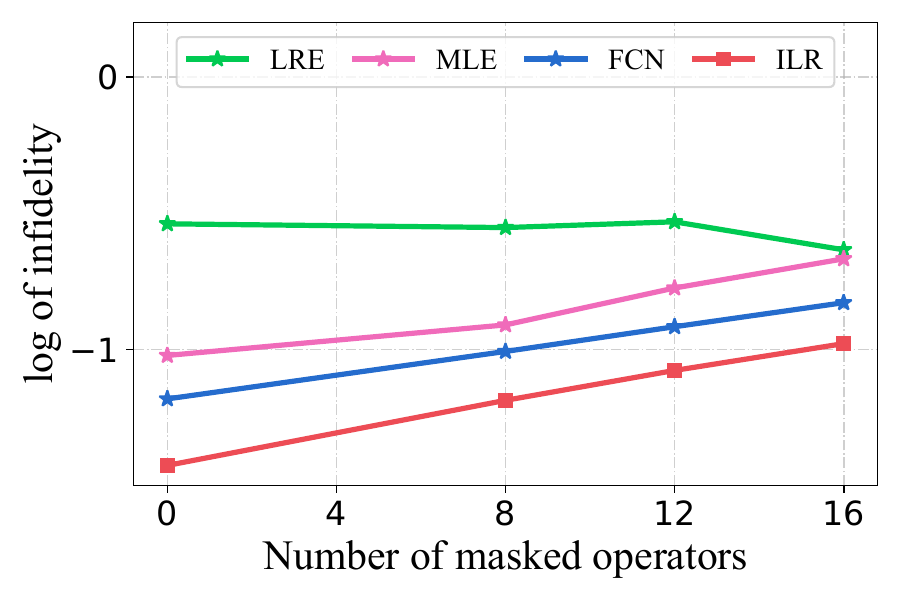}
			\centerline{(a) $N_t=10$}
			\label{mixed-N10}
		\end{minipage}
		\begin{minipage}[b]{0.32\textwidth}
			\centering
			\includegraphics[width=0.9\textwidth]{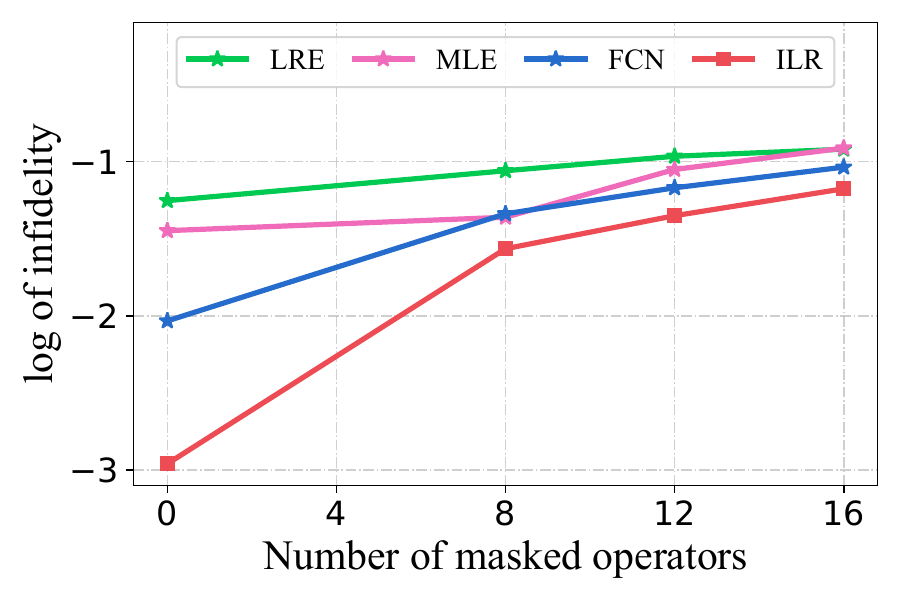}
			\centerline{(b) $N_t=100$}
			\label{mixed-N100}
		\end{minipage}
		\begin{minipage}[b]{0.32\textwidth}
			\centering
			\includegraphics[width=0.9\textwidth]{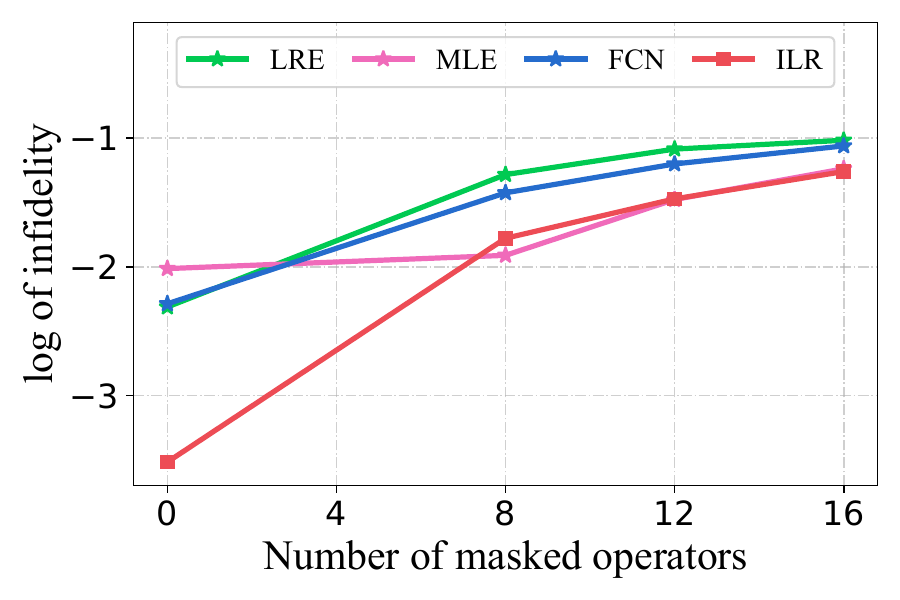}
			\centerline{(c) $N_t=1000$}
			\label{mixed-N1000}
		\end{minipage}
		\caption{Comparison results of different $N_t$ for 2-qubit mixed states}
		\label{fig:mixed_Nt}	
	\end{figure*}

The numerical results for reconstructing density matrices with different $N_t$ for pure states and mixed states are summarized in Fig.~\ref{fig:pure_Nt} and  Fig.~\ref{fig:mixed_Nt}, respectively. Under different $N_t$, the proposed approach (ILR) outperforms FCN under different cases and exhibits superiority over LRE and MLE.

To verify the function of operator embedding $\mathbf{O}$, we have introduced an experiment that excludes operator embedding to rigorously assess the performance of ILR. The results of this experiment are presented below in Table~\ref{tab:ab_oe}. As depicted in Table~\ref{tab:ab_oe}, it is clear that the performance experiences a slight decline in the absence of operator embedding, highlighting its significant contribution.

\begin{table*}[h]
	\centering
	\caption{Infidility of reconstructing density matrices for 2-qubit states with $N_t=100$, according to Fig.~\ref{fig:purestates}. ILR w/o OE means ILR method without operator embedding. }\label{tab:ab_oe}
	\setlength{\tabcolsep}{1.0mm}{
			\begin{tabular}{c|cccc|cccc}
				\toprule[1pt]
				& \multicolumn{4}{c|}{2-qubit pure states} & \multicolumn{4}{c}{4-qubit pure states} \\
				\midrule[1pt]
				Masked Number/Ratio & 0        & 8        & 12      & 16      & 0        & 20\%     & 30\%    & 40\%    \\
				\midrule[1pt]
				ILR                 & 1.85E-4  & 3.36E-4  & 8.52E-4 & 2.19E-03 & 1.28E-2  & 1.74E-2  & 2.59E-2 & 3.44E-2 \\
				ILR w/o OE          & 1.98E-4  & 5.56E-4  & 1.03E-03 & 4.19E-03 & 1.32E-2  & 2.56E-2  & 3.78E-2 & 4.78E-02 \\
				\bottomrule[1pt]
	\end{tabular}}
\end{table*}

\subsection{Additional results of predicting properties }\label{app:predictingproperties}

\begin{figure*}[h]
	\centering
	\includegraphics[width=0.32\linewidth]{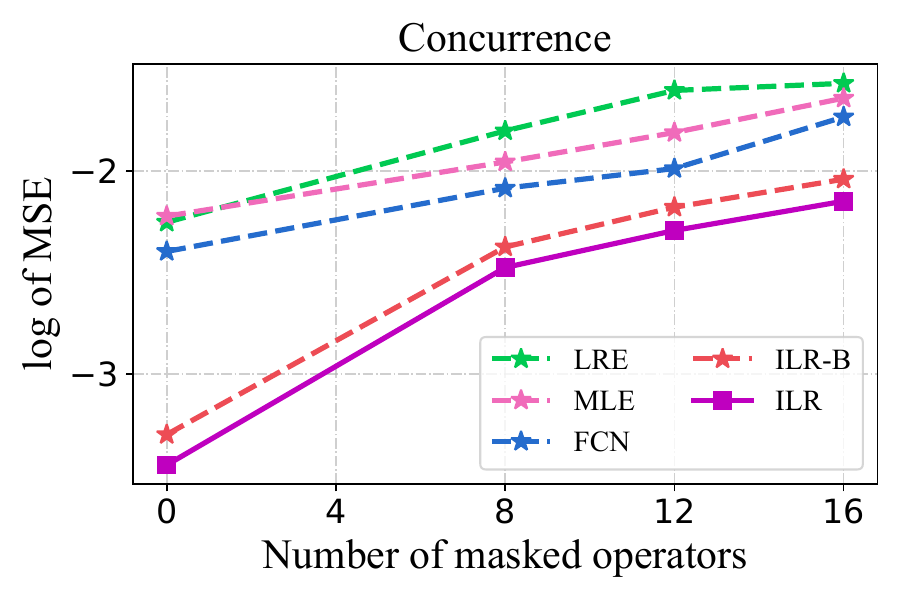}\label{sfig:mp11}
	\includegraphics[width=0.32\linewidth]{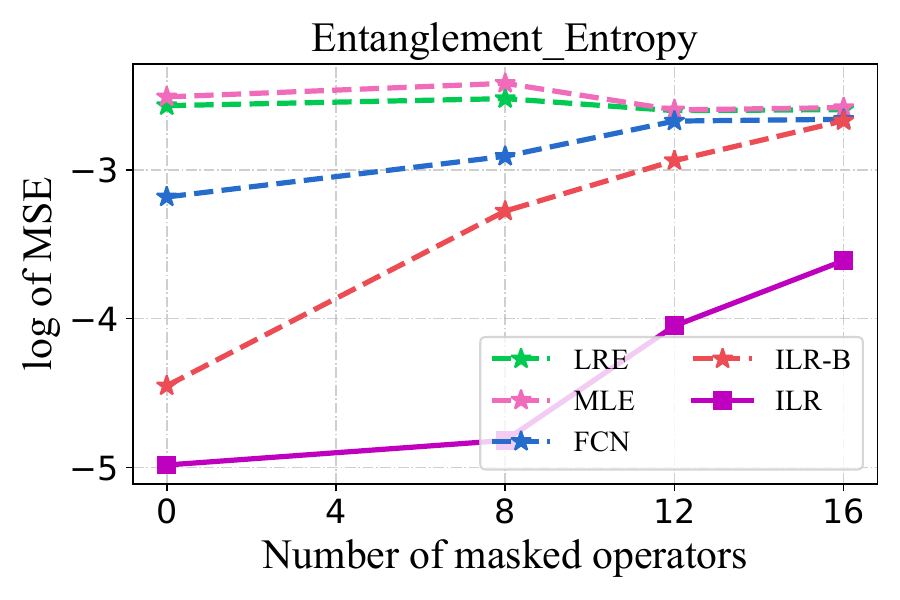}\label{sfig:mp12}
	\includegraphics[width=0.32\linewidth]{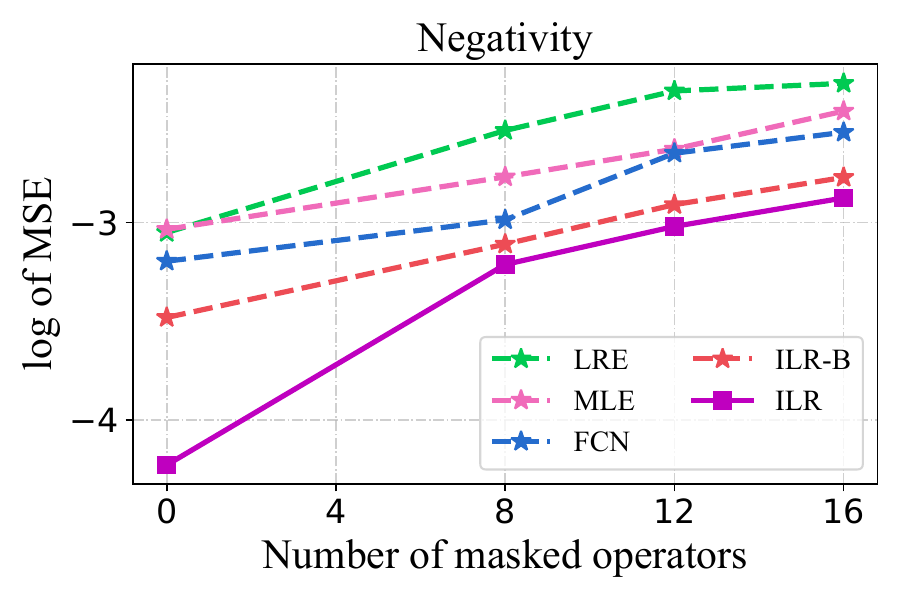}\label{sfig:mp13}
	\includegraphics[width=0.32\linewidth]{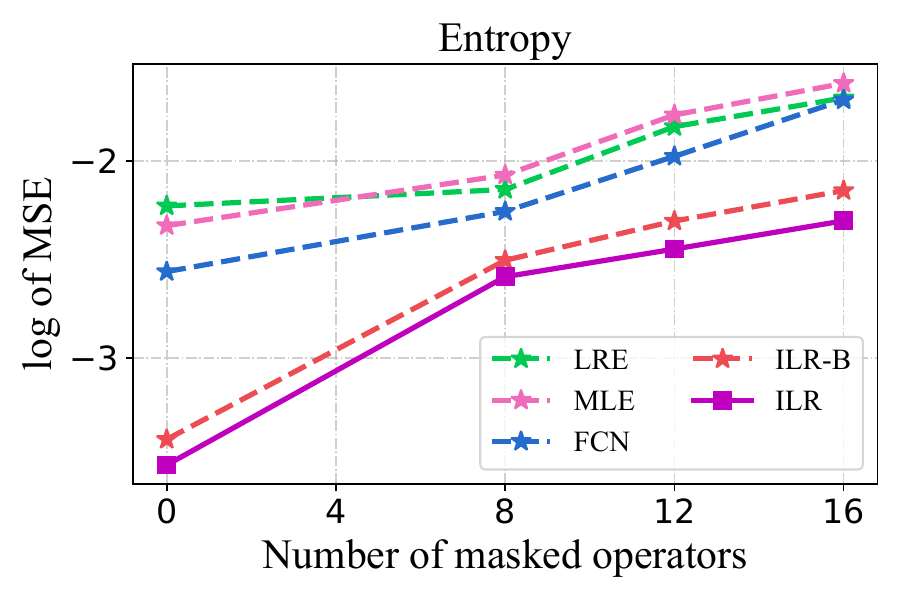}\label{sfig:mp21}
	\includegraphics[width=0.32\linewidth]{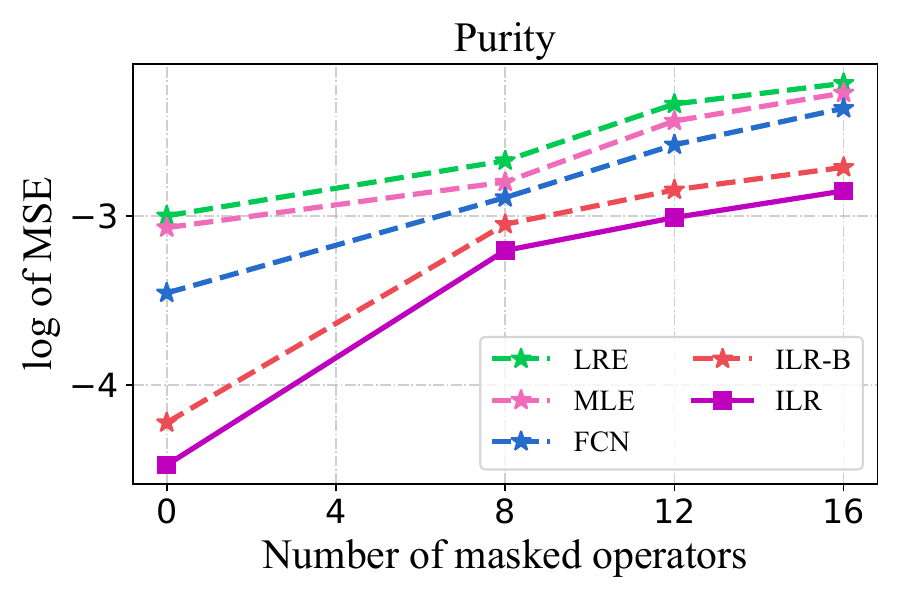}\label{sfig:mp22}
	\includegraphics[width=0.32\linewidth]{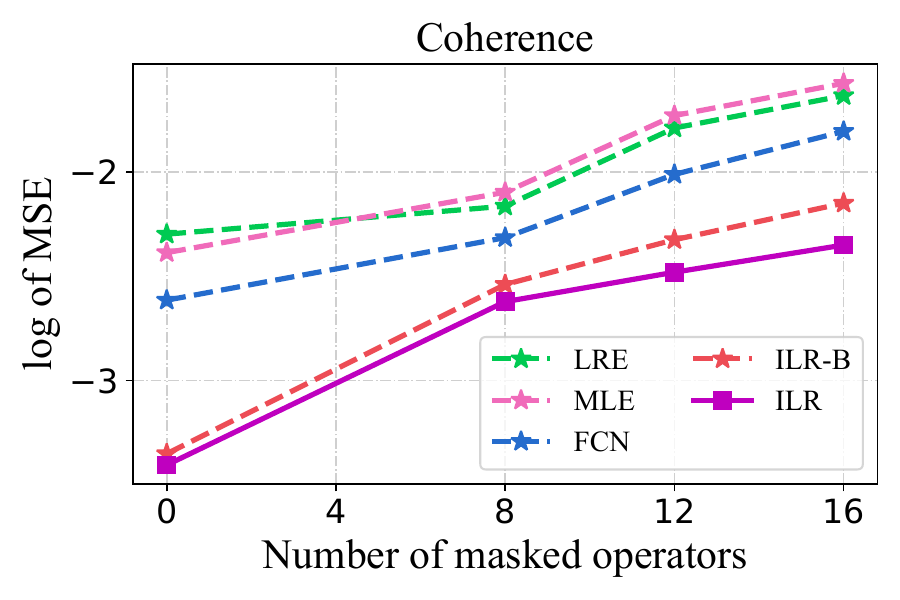}\label{sfig:mp23}
	\caption{Comparison results of predicting properties on 2-qubit mixed states ($N_t=100$). In ILR-B, FCN, MLE, and LRE, quantum properties are calculated from the reconstructed density matrices.}
	\label{fig:addition-mixed-unified}	
\end{figure*}

\textbf{2-qubit mixed states.} We focus on 6 properties of mixed states, with results summarized in Fig.~\ref{fig:addition-mixed-unified}. Clearly, the proposed ILR directly predicts properties with the highest accuracy, with great superiority over other indirect methods (FCN, MLE, LRE) that recover the density matrix using traditional algorithms and then calculate properties. The proposed ILR method also surpasses the baseline (ILR-B). 

\section{Detailed comparsion of ILR and ILR-B }\label{app:quantum-2bit-nt1000}

\begin{figure}[h]
	\centering	
	\includegraphics[width=0.9\linewidth]{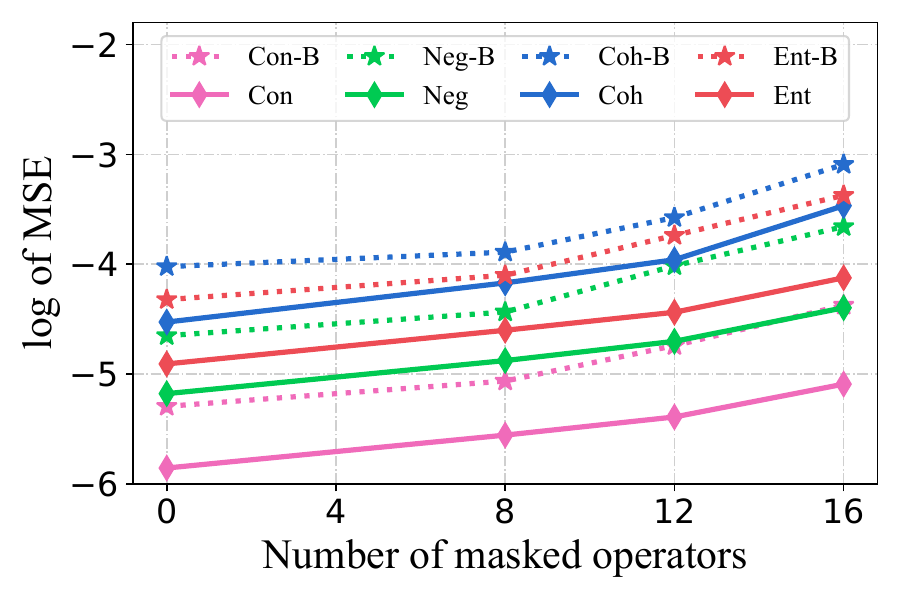}
	\caption{Two ways of property prediction for 2-qubit pure states using ILR ($N_t=1000$). The dashed lines represent that properties are calculated from the reconstructed density matrices, while the solid lines represent that properties are directly predicted from the measured frequencies.}
	\label{fig:property-pure-nt1000}	
\end{figure}

\begin{figure*}[h]
	\centering	
	{\includegraphics[width=0.45\linewidth]{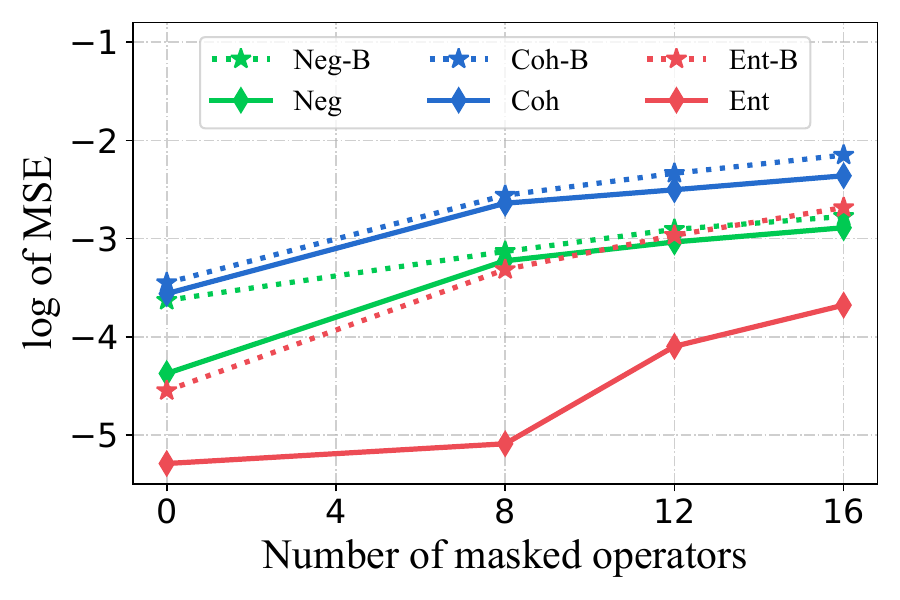}\label{sfig:mp1-nt1000}}
	{\includegraphics[width=0.45\linewidth]{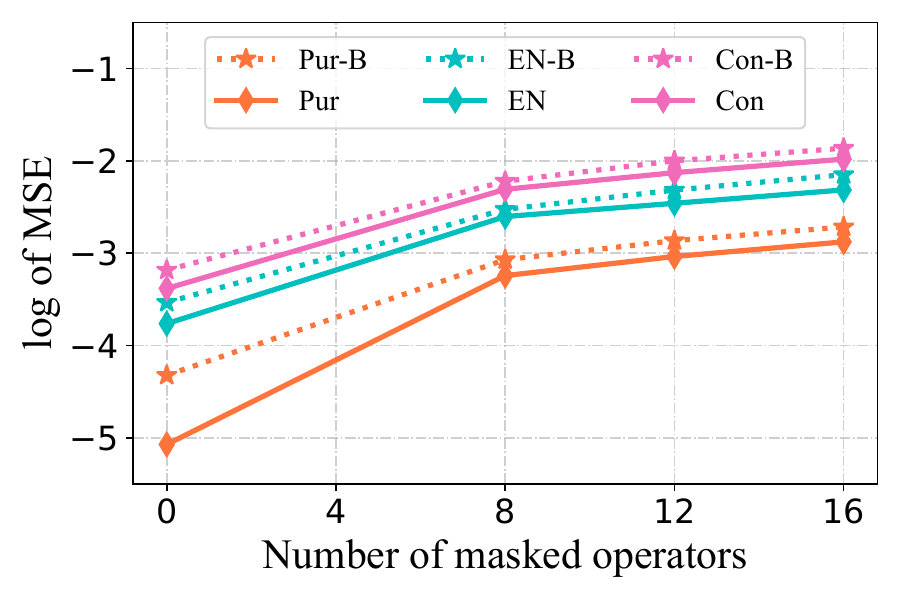}\label{sfig:mp2-nt1000}}
\caption{Two ways of property prediction for 2-qubit mixed states using ILR ($N_t=1000$). The dashed lines represent that quantum properties are calculated from the reconstructed density matrices, while the solid lines represent that properties are predicted from the imperfect measured frequencies.}
\label{fig:property-mixed-nt1000}	
\end{figure*}

Here, we compare the performance of predicting properties using ILR and the baseline that first reconstructs density matrices using ILR and then calculates properties (abbreviated as a suffix \textbf{``-B"} in the following figures). Quantum properties are abbreviated in the following figures: quantum coherence abbreviated as \textbf{``Coh"}, entanglement entropy abbreviated as \textbf{``Ent"}, negativity abbreviated as \textbf{``Neg"},  quantum entropy abbreviated as \textbf{``En"}, quantum purity abbreviated as \textbf{``Pur"} and concurrence abbreviated as \textbf{``Con"}. For example, \textbf{``Coh-B"} represents the coherence value obtained through a calculation of density matrices that are obtained from ILR.

The comparsion performance of predicting properties for pure states is summarized in Fig.~\ref{fig:property-pure-nt1000}, where predicting properties without knowing their density matrices beats the baseline results under various scenarios. Similar results have been observed for predicting 6 properties of mixed states, as shown in Fig.~\ref{fig:property-mixed-nt1000}. A closer examination of the sub-figures reveals that the gaps of entanglement entropy between ILR and ILR-B are the most significant, indicating the effectiveness of the proposed method in predicting quantum entanglement based on imperfect measured data.

\subsection{Random experiments}\label{app:papacom}

\begin{table*}[h]
	\centering
	\caption{Recovered infidelity on the 2-qubit pure states of ILR with random seeds.}\label{tab:seeds}
	\vspace{0.3cm}
	\begin{tabular}{ccccccc}
		\toprule[1pt]
		\multirow{2}{*}{Fixed}  & Seed       & 1       & 10      & 20      & 30      & 40      \\
		& Infidelity & 3.49E-4 & 3.84E-4 & 3.51E-4 & 3.51E-4 & 3.78E-4 \\
		\midrule[1pt]
		\multirow{2}{*}{Random} & Seed       & 130     & 320     & 510     & 840     & 1011    \\
		& Infidelity & 3.15E-4 & 3.61E-4 & 3.46E-4 & 3.20E-4 & 3.38E-4\\
		\bottomrule[1pt]
	\end{tabular}
\end{table*}

Table \ref{tab:seeds} presents the infidelities of our method for recovering 2-qubit pure states, evaluated under different random seed settings with fewer training epochs (100) than that in the main text (500). Two scenarios are considered: fixed seed experiments and random seed experiments. For the fixed scenario, we consistently use seeds 1, 10, 20, 30, and 40, while in the random seed scenario, the seeds are randomly selected which varies substantially from 130 to 1011. The infidelities across both scenarios are in the order of $1E{-4}$, indicating a high fidelity in the recovered states. In the fixed seed scenario, the infidelity ranges from 3.49E-4 to 3.84E-4. In the random seed scenario, despite the larger variation in seed values, the infidelity values are comparably low, ranging from 3.15E-4 to 3.61E-4. These results suggest that our method demonstrates robust performance, maintaining high fidelity in state recovery regardless of the specific seed value used. It is worth noting that apart from the experiments presented in Table \ref{tab:seeds}, we consistently set the random seed to 1 in all other experiments in this study.

\subsection{Invesigations of training and testing process}\label{app:ratio}

    \begin{table}
        \centering
        \caption{Infidelity v.s. Split ratio}
        \begin{tabular}{ccccc}
        \hline
        Ratio & 7:3 & 8:2 & 9:1 & 19:1 \\
        Infidelity & 1.75E-4 & 1.79E-4 & 1.98E-4 & 1.85E-4\\
        \hline
        \end{tabular}
        \label{tab:ratio}
        \end{table}
        
        \begin{figure}
	\centering
	\includegraphics[width=0.8\linewidth]{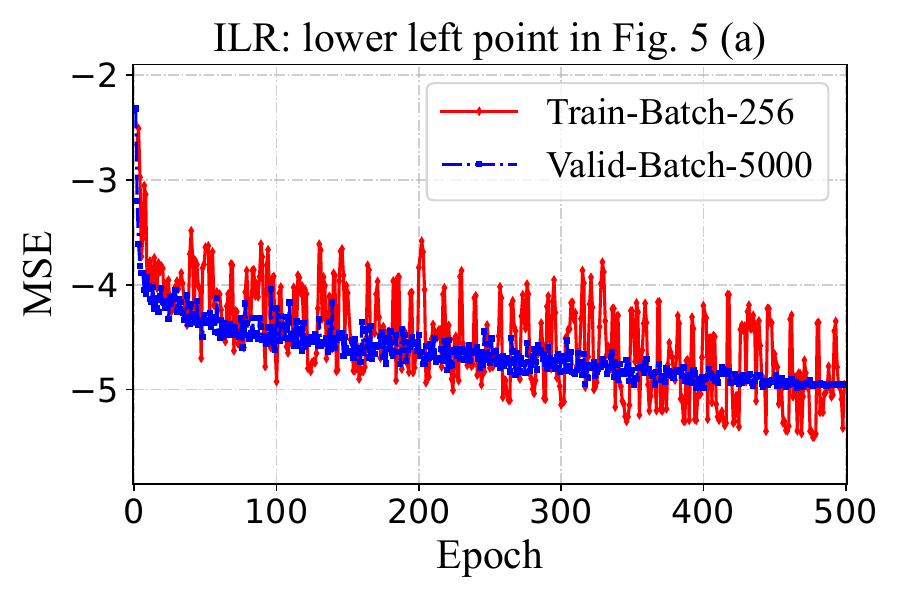}
	\caption{Training and validation curves with cosine distance metric show the training process of the lower left point in Fig. 5 (a) of the main paper. The training points are computed with a batch size of 256, which may cause some fluctuation. In contrast, the validation points are computed across the entire test dataset of 5000 samples, providing an accumulated smoothing effect.}
 \label{fig:training-curve}
\end{figure}

       \begin{table}
                \centering
                  \caption{Performance of different params on 2-qubit pure states.}
                \begin{tabular}{cccccc}
                    \hline
                    Param & 55K & 109K & 218K & 872K & 3488K \\
                    Infidelity & 3.65E-04 & 2.61E-04& 1.85E-04  & 2.53E-04 & 3.23E-04\\
                    \hline
                \end{tabular}
                \label{tab:model}
            \end{table}

            Here, we conduct additional experiments using various train/test splits (e.g., 8:2, 7:3, 9:1), with results provided in Table~\ref{tab:ratio}. These experiments consistently demonstrated that our model’s performance is stable across different test set sizes, underscoring that a smaller test set adequately captures the model’s generalization ability. Additionally, we presented the convergence curves using a 19:1 train/validation split. The curves in Fig.~\ref{fig:training-curve} demonstrate that both the training and validation sets maintain a consistent convergence trend, with no signs of overfitting. This further supports the reliability of our chosen data split. In addition, we take 2-qubit as an example and compare the performance under different model parameters, with results provided in Table~\ref{tab:model}. As we can see, an intermediate model would be a good choice. However, it is difficult to characterize the relationship between the optimal parameters in NN models and the parameters in density matrices.

\subsection{Numerical values for Fig.~\ref{fig:ab_main}-Fig.~\ref{fig:property-pure-4qubit}}

 The numerical values of Fig.~\ref{fig:ab_main}-Fig.~\ref{fig:property-pure-4qubit} are summarized in Tabel~\ref{tabel:ab_main}-Tabel~\ref{tabel:property-pure-4qubit}.

\begin{table}[h]
\centering
\caption{Numerical values in Fig.~\ref{fig:ab_main}}
\renewcommand{\arraystretch}{1.2}
\begin{tabular}{|c|c|c|c|c|}
\hline
\textbf{Mask} & \textbf{0} & \textbf{8} & \textbf{12} & \textbf{16} \\ \hline
\multicolumn{5}{|c|}{\textbf{Trans}} \\ \hline
$N_t=10$ & 9.25E-03 & 1.77E-02 & 2.60E-02 & 4.88E-02 \\ 
$N_t=100$ & 6.87E-04 & 2.01E-03 & 3.12E-03 & 8.31E-03 \\ 
$N_t=1000$ & 2.65E-04 & 7.30E-04 & 1.80E-03 & 6.02E-03 \\ \hline
\multicolumn{5}{|c|}{\textbf{S2S}} \\ \hline
$N_t=10$ & 4.26E-03 & 8.29E-03 & 1.41E-02 & 3.14E-02 \\ 
$N_t=100$ & 1.20E-04 & 3.14E-04 & 6.72E-04 & 1.90E-03 \\ 
$N_t=1000$ & 4.17E-05 & 2.58E-04 & 5.29E-04 & 1.31E-03 \\ \hline
\multicolumn{5}{|c|}{\textbf{U2S}} \\ \hline
$N_t=10$ & 4.65E-03 & 8.69E-03 & 1.49E-02 & 3.29E-02 \\ 
$N_t=100$ & 1.85E-04 & 3.36E-04 & 8.52E-04 & 2.19E-03 \\
$N_t=1000$ & 1.37E-04 & 2.83E-04 & 6.60E-04 & 1.57E-03 \\ \hline
\multicolumn{5}{|c|}{\textbf{U2U}} \\ \hline
$N_t=10$ & 4.47E-03 & 8.81E-03 & 1.66E-02 & 3.79E-02 \\
$N_t=100$ & 2.92E-04 & 5.35E-04 & 1.10E-03 & 4.12E-03 \\
$N_t=1000$ & 2.52E-04 & 4.07E-04 & 7.86E-04 & 2.47E-03 \\
\hline
\end{tabular}
\label{tabel:ab_main}
\end{table}

\begin{table}[h]
\centering
\caption{Numerical values in Fig.~\ref{fig:purestates}}
\renewcommand{\arraystretch}{1.2}
\begin{tabular}{|c|c|c|c|c|}
\hline
\textbf{Mask} & \textbf{0} & \textbf{8} & \textbf{12} & \textbf{16} \\ \hline
\multicolumn{5}{|c|}{\textbf{2 Qubits}} \\ \hline
\textbf{LRE} & 3.54E-02 & 1.52E-01 & 2.43E-01 & 2.89E-01 \\
\textbf{MLE} & 3.56E-02 & 4.35E-02 & 8.87E-02 & 1.22E-01 \\
\textbf{CNN} & 6.88E-03 & 3.27E-02 & 5.17E-02 & 7.94E-02 \\
\textbf{GAN} & 3.46E-03 & 6.46E-03 & 8.46E-03 & 1.46E-02 \\
\textbf{FCN} & 1.80E-03 & 4.26E-03 & 6.56E-03 & 9.95E-03 \\
\textbf{ICR w/o P} & 6.87E-04 & 2.01E-03 & 3.12E-03 & 8.31E-03 \\
\textbf{ICR} & 1.85E-04 & 3.36E-04 & 8.52E-04 & 2.19E-03 \\
\hline
\multicolumn{5}{|c|}{\textbf{4 Qubits}} \\ \hline
\textbf{LRE} & 9.93E-02 & 2.28E-01 & 3.31E-01 & 3.59E-01 \\
\textbf{MLE} & 8.14E-02 & 8.59E-02 & 9.15E-02 & 9.77E-02 \\
\textbf{CNN} & 9.15E-02 & 1.43E-01 & 2.21E-01 & 3.59E-01 \\
\textbf{GAN} & 1.32E-02 & 2.56E-02 & 3.78E-02 & 4.78E-02 \\
\textbf{FCN} & 6.25E-02 & 9.93E-02 & 1.51E-01 & 2.99E-01 \\
\textbf{ICR w/o P} & 2.64E-02 & 4.87E-02 & 6.34E-02 & 7.61E-02 \\
\textbf{ICR} & 1.28E-02 & 1.74E-02 & 2.59E-02 & 3.44E-02 \\
\hline
\end{tabular}
\label{tabel:purestates}
\end{table}

\begin{table}[h]
\centering
\caption{Numerical values in Fig.~\ref{fig:property-pure-2qubit}}
\renewcommand{\arraystretch}{1.2}
\begin{tabular}{|c|c|c|c|c|}
\hline
\textbf{Mask} & \textbf{0} & \textbf{8} & \textbf{12} & \textbf{16} \\ \hline
\multicolumn{5}{|c|}{\textbf{Coherence}} \\ \hline
\textbf{LRE} & 3.07E-02 & 4.89E-02 & 2.63E-01 & 7.87E-01 \\
\textbf{MLE} & 2.21E-02 & 4.91E-03 & 7.25E-02 & 3.84E-01 \\
\textbf{FCN} & 2.45E-03 & 1.18E-03 & 2.67E-03 & 2.50E-04 \\
\textbf{ILR-B} & 2.01E-04 & 9.92E-05 & 3.81E-04 & 2.06E-05 \\
\textbf{ILR} & 6.06E-05 & 3.38E-05 & 1.56E-04 & 6.27E-06 \\
\hline
\multicolumn{5}{|c|}{\textbf{Entanglement{\_}Entropy}} \\ \hline
\textbf{LRE} & 1.70E-03 & 2.72E-03 & 1.80E-02 & 3.29E-01 \\
\textbf{MLE} & 1.32E-03 & 2.56E-03 & 1.97E-02 & 3.64E-01 \\
\textbf{FCN} & 3.45E-04 & 1.66E-04 & 4.11E-04 & 1.31E-04 \\
\textbf{ILR-B} & 4.89E-05 & 2.43E-05 & 1.45E-04 & 4.98E-06 \\
\textbf{ILR} & 1.53E-05 & 8.23E-06 & 4.49E-05 & 1.65E-06 \\
\hline
\multicolumn{5}{|c|}{\textbf{Negativity}} \\ \hline
\textbf{LRE} & 7.30E-03 & 2.91E-02 & 1.69E-01 & 6.30E-01 \\
\textbf{MLE} & 1.78E-03 & 3.41E-03 & 2.62E-02 & 3.75E-01 \\
\textbf{FCN} & 1.15E-03 & 5.55E-04 & 1.91E-03 & 2.34E-04 \\
\textbf{ILR-B} & 9.37E-05 & 4.73E-05 & 2.02E-04 & 9.60E-06 \\
\textbf{ILR} & 3.67E-05 & 1.97E-05 & 1.02E-04 & 3.86E-06 \\
\hline
\end{tabular}
\label{tabel:property-pure-2qubit}
\end{table}

\begin{table}[h]
\centering
\caption{Numerical values in Fig.~\ref{fig:property-pure-4qubit}}
\renewcommand{\arraystretch}{1.2}
\begin{tabular}{|c|c|c|c|c|}
\hline
\textbf{Mask} & \textbf{0} & \textbf{8} & \textbf{12} & \textbf{16} \\ \hline
\multicolumn{5}{|c|}{\textbf{Coherence}} \\ \hline
\textbf{LRE} &7.58E-02 & 3.60E-01 & 5.18E-01 & 7.41E-01 \\
\textbf{FCN} &1.79E-02 & 2.35E-02 & 1.83E-01 & 4.17E-01 \\
\textbf{ILR-B} &8.63E-04 & 2.15E-03 & 2.96E-03 & 5.59E-03 \\
\textbf{ILR} &2.98E-04 & 4.57E-04 & 5.78E-04 & 8.69E-04 \\
\hline
\multicolumn{5}{|c|}{\textbf{Entanglement\_Entropy}} \\ \hline
\textbf{LRE} &3.19E-03 & 1.99E-02 & 3.43E-02 & 4.45E-02 \\
\textbf{FCN} &5.55E-03 & 8.01E-03 & 2.01E-02 & 5.51E-02 \\
\textbf{ILR-B} &4.66E-04 & 1.36E-03 & 1.98E-03 & 4.27E-03 \\
\textbf{ILR} &1.23E-04 & 1.72E-04 & 2.00E-04 & 2.82E-04 \\
\hline
\multicolumn{5}{|c|}{\textbf{Negativity}} \\ \hline
\textbf{LRE} &3.09E-02 & 1.21E-01 & 2.28E-01 & 3.05E-01 \\
\textbf{FCN} &1.15E-02 & 2.26E-02 & 9.08E-02 & 1.63E-01 \\
\textbf{ILR-B} &8.35E-04 & 1.81E-03 & 2.57E-03 & 5.28E-03 \\
\textbf{ILR} &5.03E-04 & 1.23E-03 & 1.34E-03 & 1.61E-03 \\
\hline
\end{tabular}
\label{tabel:property-pure-4qubit}
\end{table}

\subsection{Algorithm description for QST}

	\begin{figure}[h]
		
			\begin{algorithm}[H]
				\small
				\caption{\small Description for QST.}\label{alg2}
				\begin{algorithmic}[1]
					\REQUIRE Operators $\mathbf{O}$, Frequencies $\mathbf{f}$, Ground-truth vectors $\bm \nu$ or $\bm \mu$, Masking list $\mathbf{m}$, Training epochs $T$, Dataset steps $S$. 
					\ENSURE Obtaining the QST model $\Phi({\Theta_{q}})$.
					\STATE Build QST model $\Phi({\Theta_{q}})$ with Encoder $\Phi({\Theta_{e}})$ and State Decoder $\Phi({\Theta_{sd}})$;
					\STATE Load $\Phi({\Theta_{e}})$ from $\Phi({\Theta_{p}})$, and initialize $\Phi({\Theta_{sd}})$ with default Kaiming initialization;
					\STATE \textbf{for} $t=1,2,\cdots,T$ \textbf{do}
					\STATE \quad \textbf{for} $s=1,2,\cdots,S$ \textbf{do}
					\STATE \qquad \textbf{if} Training is \textit{\color{blue}{Separate}} \textbf{then}
					\STATE \quad\qquad Use the fixed mask number $m$;
					\STATE \qquad \textbf{if} Training is \textit{\color{blue}{Unified}} \textbf{then}
					\STATE \quad\qquad Random select $m$ from $\mathbf{m}$;
					\STATE \qquad Forward using $\mathbf{\widetilde{f}}$ and $\mathbf{\widetilde{O}}$ obtained with $m$;
					\STATE \qquad Backward by minimizing $MSE(\bm{\hat{\nu}}, \mathbf{\bm\nu})$ or \\ \qquad$MSE(\bm{\hat{\mu}}, \mathbf{\bm\mu})$;
					\STATE \qquad Update ${\Theta_{sd}}$ with gradient descent; 
					\RETURN optimal $\Phi({\Theta_{q}})$.
				\end{algorithmic}
			\end{algorithm}
	\end{figure}

\bibliographystyle{ieeetr}
\bibliography{mybib}

\begin{thebibliography}{10}

\bibitem{gebhart2023learning}
V.~Gebhart, R.~Santagati, A.~A. Gentile, E.~M. Gauger, D.~Craig, N.~Ares,
  L.~Banchi, F.~Marquardt, L.~Pezz{\`e}, and C.~Bonato, ``Learning quantum
  systems,'' {\em Nature Reviews Physics}, pp.~1--16, 2023.

\bibitem{rambach2021robust}
M.~Rambach, M.~Qaryan, M.~Kewming, C.~Ferrie, A.~G. White, and J.~Romero,
  ``Robust and efficient high-dimensional quantum state tomography,'' {\em
  Physical Review Letters}, vol.~126, no.~10, p.~100402, 2021.

\bibitem{jevzek2003quantum}
M.~Je{\v{z}}ek, J.~Fiur{\'a}{\v{s}}ek, and Z.~Hradil, ``Quantum inference of
  states and processes,'' {\em Physical Review A}, vol.~68, no.~1, p.~012305,
  2003.

\bibitem{haah2017sample}
J.~Haah, A.~W. Harrow, Z.~Ji, X.~Wu, and N.~Yu, ``Sample-optimal tomography of
  quantum states,'' {\em IEEE Transactions on Information Theory}, vol.~63,
  no.~9, pp.~5628–--5641, 2017.

\bibitem{qin2023stable}
Z.~Qin, C.~Jameson, Z.~Gong, M.~B. Wakin, and Z.~Zhu, ``Stable tomography for
  structured quantum states,'' {\em arXiv preprint arXiv:2306.09432}, 2023.

\bibitem{len2022quantum}
Y.~L. Len, T.~Gefen, A.~Retzker, and J.~Ko{\l}ody{\'n}ski, ``Quantum metrology
  with imperfect measurements,'' {\em Nature Communications}, vol.~13, no.~1,
  p.~6971, 2022.

\bibitem{chantasri2019quantum}
A.~Chantasri, S.~Pang, T.~Chalermpusitarak, and A.~N. Jordan, ``Quantum state
  tomography with time-continuous measurements: reconstruction with resource
  limitations,'' {\em Quantum Studies: Mathematics and Foundations}, pp.~1--25,
  2019.

\bibitem{teo2012incomplete}
Y.~S. Teo, B.~Stoklasa, B.-G. Englert, J.~{\v{R}}eh{\'a}{\v{c}}ek, and
  Z.~Hradil, ``Incomplete quantum state estimation: A comprehensive study,''
  {\em Physical Review A}, vol.~85, no.~4, p.~042317, 2012.

\bibitem{torlai2018neural}
G.~Torlai, G.~Mazzola, J.~Carrasquilla, M.~Troyer, R.~Melko, and G.~Carleo,
  ``Neural-network quantum state tomography,'' {\em Nature Physics}, vol.~14,
  no.~5, pp.~447--450, 2018.

\bibitem{carrasquilla2019reconstructing}
J.~Carrasquilla, G.~Torlai, R.~G. Melko, and L.~Aolita, ``Reconstructing
  quantum states with generative models,'' {\em Nature Machine Intelligence},
  vol.~1, no.~3, pp.~155--161, 2019.

\bibitem{cha2021attention}
P.~Cha, P.~Ginsparg, F.~Wu, J.~Carrasquilla, P.~L. McMahon, and E.-A. Kim,
  ``Attention-based quantum tomography,'' {\em Machine Learning: Science and
  Technology}, vol.~3, no.~1, p.~01LT01, 2021.

\bibitem{ahmed2021quantum}
S.~Ahmed, C.~S. Mu{\~n}oz, F.~Nori, and A.~F. Kockum, ``Quantum state
  tomography with conditional generative adversarial networks,'' {\em Physical
  Review Letters}, vol.~127, no.~14, p.~140502, 2021.

\bibitem{danaci2021machine}
O.~Danaci, S.~Lohani, B.~T. Kirby, and R.~T. Glasser, ``Machine learning
  pipeline for quantum state estimation with incomplete measurements,'' {\em
  Machine Learning: Science and Technology}, vol.~2, no.~3, p.~035014, 2021.

\bibitem{transformer2017}
A.~Vaswani, N.~Shazeer, N.~Parmar, J.~Uszkoreit, L.~Jones, A.~N. Gomez,
  {\L}.~Kaiser, and I.~Polosukhin, ``Attention is all you need,'' {\em Advances
  in Neural Information Processing Systems}, vol.~30, 2017.

\bibitem{devlin2018bert}
J.~Devlin, M.-W. Chang, K.~Lee, and K.~Toutanova, ``Bert: Pre-training of deep
  bidirectional transformers for language understanding,'' {\em arXiv preprint
  arXiv:1810.04805}, 2018.

\bibitem{qi2013quantum}
B.~Qi, Z.~Hou, L.~Li, D.~Dong, G.~Xiang, and G.~Guo, ``Quantum state tomography
  via linear regression estimation,'' {\em Scientific Reports}, vol.~3,
  p.~3496, 2013.

\bibitem{opatrny1997least}
T.~Opatrn{\`y}, D.-G. Welsch, and W.~Vogel, ``Least-squares inversion for
  density-matrix reconstruction,'' {\em Physical Review A}, vol.~56, no.~3,
  p.~1788, 1997.

\bibitem{qi2017adaptive}
B.~Qi, Z.~Hou, Y.~Wang, D.~Dong, H.-S. Zhong, L.~Li, G.-Y. Xiang, H.~M.
  Wiseman, C.-F. Li, and G.-C. Guo, ``Adaptive quantum state tomography via
  linear regression estimation: Theory and two-qubit experiment,'' {\em npj
  Quantum Information}, vol.~3, p.~19, 2017.

\bibitem{huszar2012adaptive}
F.~Husz{\'a}r and N.~M. Houlsby, ``Adaptive {B}ayesian quantum tomography,''
  {\em Physical Review A}, vol.~85, no.~5, p.~052120, 2012.

\bibitem{flammia2012quantum}
S.~T. Flammia, D.~Gross, Y.-K. Liu, and J.~Eisert, ``Quantum tomography via
  compressed sensing: error bounds, sample complexity and efficient
  estimators,'' {\em New Journal of Physics}, vol.~14, no.~9, p.~095022, 2012.

\bibitem{huang2020predicting}
H.-Y. Huang, R.~Kueng, and J.~Preskill, ``Predicting many properties of a
  quantum system from very few measurements,'' {\em Nature Physics}, vol.~16,
  no.~10, pp.~1050--1057, 2020.

\bibitem{magesan2012characterizing}
E.~Magesan, J.~M. Gambetta, and J.~Emerson, ``Characterizing quantum gates via
  randomized benchmarking,'' {\em Physical Review A—Atomic, Molecular, and
  Optical Physics}, vol.~85, no.~4, p.~042311, 2012.

\bibitem{xu2018neural}
Q.~Xu and S.~Xu, ``Neural network state estimation for full quantum state
  tomography,'' {\em arXiv preprint arXiv:1811.06654}, 2018.

\bibitem{palmieri2020experimental}
A.~M. Palmieri, E.~Kovlakov, F.~Bianchi, D.~Yudin, S.~Straupe, J.~D. Biamonte,
  and S.~Kulik, ``Experimental neural network enhanced quantum tomography,''
  {\em npj Quantum Information}, vol.~6, p.~20, 2020.

\bibitem{lohani2020machine}
S.~Lohani, B.~Kirby, M.~Brodsky, O.~Danaci, and R.~T. Glasser, ``Machine
  learning assisted quantum state estimation,'' {\em Machine Learning: Science
  and Technology}, vol.~1, no.~3, p.~035007, 2020.

\bibitem{lohani2021experimental}
S.~Lohani, T.~A. Searles, B.~T. Kirby, and R.~T. Glasser, ``On the experimental
  feasibility of quantum state reconstruction via machine learning,'' {\em IEEE
  Transactions on Quantum Engineering}, vol.~2, pp.~1--10, 2021.

\bibitem{zhang2022transformer}
Y.-H. Zhang and M.~Di~Ventra, ``Transformer quantum state: A multi-purpose
  model for quantum many-body problems,'' {\em arXiv preprint
  arXiv:2208.01758}, 2022.

\bibitem{zhong2022quantum}
L.~Zhong, C.~Guo, and X.~Wang, ``Quantum state tomography inspired by language
  modeling,'' {\em arXiv preprint arXiv:2212.04940}, 2022.

\bibitem{zhu2022flexible}
Y.~Zhu, Y.-D. Wu, G.~Bai, D.-S. Wang, Y.~Wang, and G.~Chiribella, ``Flexible
  learning of quantum states with generative query neural networks,'' {\em
  Nature communications}, vol.~13, no.~1, pp.~1--10, 2022.

\bibitem{carleo2017solving}
G.~Carleo and M.~Troyer, ``Solving the quantum many-body problem with
  artificial neural networks,'' {\em Science}, vol.~355, no.~6325,
  pp.~602--606, 2017.

\bibitem{schmale2022efficient}
T.~Schmale, M.~Reh, and M.~G{\"a}rttner, ``Efficient quantum state tomography
  with convolutional neural networks,'' {\em npj Quantum Information}, vol.~8,
  p.~15, 2022.

\bibitem{hinton1993autoencoders}
G.~E. Hinton and R.~Zemel, ``Autoencoders, minimum description length and
  helmholtz free energy,'' {\em Advances in Neural Information Processing
  Systems}, vol.~6, 1993.

\bibitem{kingma2013auto}
D.~P. Kingma and M.~Welling, ``Auto-encoding variational {B}ayes,'' {\em arXiv
  preprint arXiv:1312.6114}, 2013.

\bibitem{ho2020denoising}
J.~Ho, A.~Jain, and P.~Abbeel, ``Denoising diffusion probabilistic models,''
  {\em Advances in Neural Information Processing Systems}, vol.~33,
  pp.~6840--6851, 2020.

\bibitem{dong2015image}
C.~Dong, C.~C. Loy, K.~He, and X.~Tang, ``Image super-resolution using deep
  convolutional networks,'' {\em IEEE Transactions on Pattern Analysis and
  Machine Intelligence}, vol.~38, no.~2, pp.~295--307, 2015.

\bibitem{cambria2014jumping}
E.~Cambria and B.~White, ``Jumping nlp curves: A review of natural language
  processing research,'' {\em IEEE Computational Intelligence Magazine},
  vol.~9, no.~2, pp.~48--57, 2014.

\bibitem{raffel2020exploring}
C.~Raffel, N.~Shazeer, A.~Roberts, K.~Lee, S.~Narang, M.~Matena, Y.~Zhou,
  W.~Li, and P.~J. Liu, ``Exploring the limits of transfer learning with a
  unified text-to-text transformer,'' {\em The Journal of Machine Learning
  Research}, vol.~21, no.~1, pp.~5485--5551, 2020.

\bibitem{radford2018improving}
A.~Radford, K.~Narasimhan, T.~Salimans, I.~Sutskever, {\em et~al.}, ``Improving
  language understanding by generative pre-training,'' 2018.

\bibitem{alexey2020image}
A.~Dosovitskiy, L.~Beyer, A.~Kolesnikov, D.~Weissenborn, X.~Zhai,
  T.~Unterthiner, M.~Dehghani, M.~Minderer, G.~Heigold, S.~Gelly, {\em et~al.},
  ``An image is worth 16x16 words: Transformers for image recognition at
  scale,'' {\em arXiv preprint arXiv:2010.11929}, 2020.

\bibitem{liu2021swin}
Z.~Liu, Y.~Lin, Y.~Cao, H.~Hu, Y.~Wei, Z.~Zhang, S.~Lin, and B.~Guo, ``Swin
  transformer: Hierarchical vision transformer using shifted windows,'' in {\em
  Proceedings of the IEEE/CVF International Conference on Computer Vision},
  pp.~10012--10022, 2021.

\bibitem{he2022masked}
K.~He, X.~Chen, S.~Xie, Y.~Li, P.~Doll{\'a}r, and R.~Girshick, ``Masked
  autoencoders are scalable vision learners,'' in {\em Proceedings of the
  IEEE/CVF Conference on Computer Vision and Pattern Recognition},
  pp.~16000--16009, 2022.

\bibitem{lecun1998lenet-5}
Y.~LeCun, L.~Bottou, Y.~Bengio, and P.~Haffner, ``Gradient-based learning
  applied to document recognition,'' {\em Proceedings of the IEEE}, vol.~86,
  no.~11, pp.~2278--2324, 1998.

\bibitem{brown2020language}
T.~Brown, B.~Mann, N.~Ryder, M.~Subbiah, J.~D. Kaplan, P.~Dhariwal,
  A.~Neelakantan, P.~Shyam, G.~Sastry, A.~Askell, {\em et~al.}, ``Language
  models are few-shot learners,'' {\em Advances in Neural Information
  Processing Systems}, vol.~33, pp.~1877--1901, 2020.

\bibitem{nielsen2010quantum}
M.~A. Nielsen and I.~L. Chuang, {\em Quantum Computation and Quantum
  Information}.
\newblock Cambridge University Press, 2010.

\bibitem{liu2004quantum}
Y.-x. Liu, L.~Wei, and F.~Nori, ``Quantum tomography for solid-state qubits,''
  {\em EPL (Europhysics Letters)}, vol.~67, no.~6, p.~874, 2004.

\bibitem{ma2021comparative}
H.~Ma, D.~Dong, I.~R. Petersen, C.-J. Huang, and G.-Y. Xiang, ``A comparative
  study on how neural networks enhance quantum state tomography,'' {\em arXiv
  preprint arXiv:2111.09504}, 2021.

\bibitem{higham1990analysis}
N.~J. Higham, {\em Analysis of the Cholesky Decomposition of a Semi-definite
  Matrix}.
\newblock Oxford University Press, 1990.

\bibitem{mezzadri2006generate}
F.~Mezzadri, ``How to generate random matrices from the classical compact
  groups,'' {\em arXiv preprint math-ph/0609050}, 2006.

\bibitem{forrester2007eigenvalue}
P.~J. Forrester and T.~Nagao, ``Eigenvalue statistics of the real ginibre
  ensemble,'' {\em Physical Review Letters}, vol.~99, no.~5, p.~050603, 2007.

\bibitem{ozawa2000entanglement}
M.~Ozawa, ``Entanglement measures and the {H}ilbert--{S}chmidt distance,'' {\em
  Physics Letters A}, vol.~268, no.~3, pp.~158--160, 2000.

\bibitem{de2008choice}
M.~D. De~Burgh, N.~K. Langford, A.~C. Doherty, and A.~Gilchrist, ``Choice of
  measurement sets in qubit tomography,'' {\em Physical Review A}, vol.~78,
  no.~5, p.~052122, 2008.

\bibitem{adamson2010improving}
R.~Adamson and A.~M. Steinberg, ``Improving quantum state estimation with
  mutually unbiased bases,'' {\em Physical Review Letters}, vol.~105, no.~3,
  p.~030406, 2010.

\bibitem{koutny2022neural}
D.~Koutn{\`y}, L.~Motka, Z.~Hradil, J.~{\v{R}}eh{\'a}{\v{c}}ek, and L.~L.
  S{\'a}nchez-Soto, ``Neural-network quantum state tomography,'' {\em Physical
  Review A}, vol.~106, no.~1, p.~012409, 2022.

\bibitem{rambach2022efficient}
M.~Rambach, A.~Youssry, M.~Tomamichel, and J.~Romero, ``Efficient quantum state
  tracking in noisy environments,'' {\em Quantum Science and Technology},
  vol.~8, no.~1, p.~015010, 2022.

\bibitem{ivanova2023optimal}
V.~N. Ivanova-Rohling, N.~Rohling, and G.~Burkard, ``Optimal quantum state
  tomography with noisy gates,'' {\em EPJ Quantum Technology}, vol.~10, p.~25,
  2023.

\bibitem{baumgratz2014quantifying}
T.~Baumgratz, M.~Cramer, and M.~B. Plenio, ``Quantifying coherence,'' {\em
  Physical Review Letters}, vol.~113, no.~14, p.~140401, 2014.

\end{thebibliography}

\end{document}